\documentclass{aa}  
\usepackage{graphicx}
\usepackage[colorlinks=true,linkcolor=blue,citecolor=blue]{hyperref}
\usepackage{txfonts}
\usepackage{orcidlink}
\usepackage[noabbrev, nameinlink]{cleveref} 
\hypersetup{
    colorlinks=true,
    linkcolor=blue,
    filecolor=magenta,      
    urlcolor=blue,
    pdftitle={Overleaf Example},
    pdfpagemode=FullScreen,
    }
\Crefname{section}{Sect.}{Sects.}
\Crefname{figure}{Fig.}{Figs.}
\Crefname{equation}{Eq.}{Eqs.}

\let\em\it
\usepackage{amstext}

\begin{document} 
         
\title{Impact of hot exozodiacal dust on the polarimetric\\ analysis of close-in exoplanets}
        
        
\author{K. Ollmann\ \orcidlink{0009-0003-6954-5252}
\and S. Wolf\ \orcidlink{0000-0001-7841-3452}
\and M. Lietzow\ \orcidlink{0000-0001-9511-3371}
\and T. A. Stuber\ \orcidlink{0000-0003-2185-0525}}
                
\institute{Institute of Theoretical Physics and Astrophysics, Kiel University, Leibnizstr. 15, 24118 Kiel,\\ Germany\\
\email{kollmann@astrophysik.uni-kiel.de}
}
        \date{Received 31 May 2023/ Accepted 26 July 2023}

    \abstract
{Hot exozodiacal dust (HEZD) found around main-sequence stars through interferometric observations in the photometric bands \(H\) to \(L\) is located close to the dust sublimation radius, potentially at orbital radii comparable to those of close-in exoplanets. Consequently, HEZD has a potential influence on the analysis of the scattered-light polarization of close-in exoplanets and vice versa.}
{We analyze the impact of HEZD on the polarimetric characterization of close-in exoplanets. This study is motivated in particular by the recently proven feasibility of exoplanet polarimetry.}
{Applying the 3D Monte Carlo radiative transfer code POLARIS in an extended and optimized version for radiative transfer in exoplanetary atmospheres and an analytical tool for modeling the HEZD, we simulated and compared the polarization characteristics of the wavelength-dependent scattered-light polarization of HEZD and close-in exoplanets. As a starting point for our analysis, we defined a reference model consisting of a close-in exoplanet with a scattered-light polarization consistent with the upper limit determined for WASP-18b, and a HEZD consistent with the near-infrared excess detected for HD 22484 (10 Tau).}
{The varied parameters are the planetary phase angle (\(0^\circ-180^\circ\)), the dust grain radius (\(0.02\ \mu\)m \(- \ 10\ \mu\)m), the HEZD mass (\(10^{-10}\)\(\rm{M}_{\oplus}\)  \(-\ 10^{-8}\)\(\rm{M}_{\oplus}\)), the orbital inclination (\(0^\circ-90^\circ\)), the composition of the planetary atmosphere (Mie and Rayleigh scattering atmosphere), the orbital radius of HEZD (\(0.02\) au  \(-\  0.4\) au), and the planetary orbital radius (\(0.01\) au  \(-\  0.05\) au). The dust grain radius has the strongest influence on the polarimetric analysis due to its significant impact on the wavelength-dependent polarization characteristics and the total order of magnitude of the scattered-light polarization. In certain scenarios, the scattered-light polarization of the HEZD even exceeds that of the close-in exoplanet, for example for a dust grain radius of \(0.1\ \mu\)m, a HEZD mass of \(8\times 10^{-10} \rm{M}_{\oplus}\), an orbital radius of HEZD of \(0.04\) au and an orbital inclination of \(90^\circ\).}
{The presence of HEZD potentially has a significant impact on the polarimetric investigations of close-in exoplanets. Furthermore, interferometric observations are required to better constrain the parameter space for HEZD and thus the possible resulting scattered-light polarization.}
        
\titlerunning{Impact of HEZD on the polarimetric analysis of close-in exoplanets}
\authorrunning{Ollmann et al.}
\maketitle
        

\section{Introduction}
A strong near-infrared (NIR) excess is detected at the \(1\%\) level around more than two dozen main-sequence stars by interferometric observations in the photometric bands \(H\) to \(L\)  (e.g., \citealt{Absil, Absil2013}; \citealt{Ertel, Ertel2016}). This excess is attributed to submicrometer dust grains in close vicinity to stars (e.g., \citealt{Di Folco}) and is often referred to as hot exozodi or hot exozodiacal dust (hereafter: HEZD). A simple but so far sufficient model for fitting the observed NIR excesses is an optically thin, geometrically narrow dust ring located close to the dust sublimation radius. Graphite grains with radii often smaller than their blowout size, a steep particle distribution and low HEZD masses around \(10^{-10}\) \(\rm{M}_{\oplus}\) to  \(10^{-8}\)  \(\rm{M}_{\oplus}\) are suitable for fitting the observed NIR excess (e.g., \citealt{Kirchschlager2018}). Because of the high contrast and the small angular distance between this circumstellar dust and the star, observations of HEZD were so far only possible through NIR and mid-infrared (MIR) long-baseline and nulling interferometry (e.g., \citealt{Absil2021}). Moreover, no clear correlations between the NIR excess and MIR or far-infrared (FIR) excess tracing dust farther out have been observed so far (\citealt{Millan-Gabet}; \citealt{Mennesson}). In addition, the contribution of silicate grains is negligible because the weak MIR excess is inconsistent with the otherwise expected strong \(N\) band feature (\citealt{Akeson}; \citealt{Kirchschlager2017}).\\
\\
The mechanisms that sustain HEZD at the observed level are currently unclear because these small grains are thought to sublimate rapidly or be blown out of the system (\citealt{Backman}; \citealt{Wyatt2007}; \citealt{Lebreton}). The replenishment of dust grains \textup{in situ} through a steady-state collisional cascade close to the star can most likely also be excluded because the dust lifetime at these distances is short (e.g., \citealt{Wyatt}; \citealt{Lebreton}). Several alternative scenarios to explain the presence of HEZD, for instance magnetic trapping or cometary supply, have been discussed, but none of these scenarios individually or in combination provided a comprehensive explanation for the transport of the dust to the inner regions nor how it can survive there or can be replenished efficiently (\citealt{Kobayashi};  \citealt{Faramaz}; \citealt{Sezestre}; \citealt{Stamm}; \citealt{Kimura}; \citealt{Rigley}; \citealt{Pearce2020, Pearce}).\\
This HEZD may offer a way to probe the inner regions of extrasolar planetary systems because its presence might constrain the architecture of the planetary system (e.g., \citealt{Kral}).
Direct imaging of those close-in planets is not yet possible because the faint planetary signal is lost in the bright stellar glare. Thus, polarimetry has become a useful tool in recent years to potentially distinguish and characterize the weak, polarized signal that is reflected by the planet from the direct stellar radiation (e.g., \citealt{Bott2016}). Close-in planets such as hot Jupiters are well suited for polarimetric investigations considering their large radii and the proximity to their host star and thus the high fraction of radiation that is scattered off their atmosphere (\citealt{Seager}; \citealt{Stam2004}). High-precision polarimeters, such as the ground-based HIgh-Precision Polarimetric Instruments (HIPPI, HIPPI-2; \citealt{Bailey2015,Bailey2020}) and the POlarimeter at Lick for Inclination Studies of Hot jupiters 2 (POLISH2; \citealt{Polish2}) allow measuring the polarized flux of several hot Jupiters, for example, within the exoplanet detection program –"Wide Angle Search for Planets"– (WASP) for the hot Jupiter WASP-18b at the parts per million (ppm) level (\citealt{Bott}; \citealt{Bailey2021}).\\ At the same time the possible presence of small hot dust grains potentially interferes with the polarimetric characterization of planets in the habitable zone (\citealt{Agol2007}; \citealt{Beckwith2008}; \citealt{Roberge}; \citealt{Stapelfeldt}). Therefore, HEZD has to be considered in the attempt to characterize exoplanets via their scattered-light polarization.\\
\\ 
We investigate the influence of HEZD on the analysis of the wavelength-dependent scattered-light polarization of close-in exoplanets. For this purpose we investigate characteristic features of the individual signatures of a close-in exoplanet and a HEZD on the wavelength-dependent polarization signal in the wavelength range from 550 nm to \(4\ \mu\)m, covering the photometric bands and the respective representative wavelengths \(V (\mathrm{550\ nm}),\  R (\mathrm{650\ nm}),\  Z (\mathrm{878\ nm}),\  I (1.00\ \mu\mathrm{m}),\  J (1.25\ \mu\mathrm{m}),\\  H (1.65\ \mu\mathrm{m}),\  K (2.22\ \mu\mathrm{m})\) and \(L (3.45\ \mu\mathrm{m})\).\\
This article is organized as follows: In \Cref{reference} we give a brief overview about the assumptions based on which we simulate the scattered-light polarization. Moreover, we describe the reference model that provides the basis for the analysis of the influence of selected physical parameters on the polarization characteristics of the close-in planet and the HEZD. In \Cref{results1} we study the impact of selected model parameters on the wavelength-dependent polarization degree (\Cref{wavelength-dependence}) and on the polarized flux intensity (\Cref{V-band}) of the HEZD and the close-in planet. Moreover, the polarization degrees are discussed in the context of the intrinsic stellar polarization, especially for observed active stars. Our findings are summarized in \Cref{Summary}.
\section{Reference model}\label{reference}
In \Cref{Linear polarization degree} we define the polarimetric quantities that we used for the qualitative description of the linear polarization. Subsequently, we define a reference model consisting of these components, that is a star (\Cref{Reference star}), a close-in planet (\Cref{Reference close-in planet}) and a HEZD (\Cref{Reference hot exozodiacal dust}).

\subsection{Polarimetric quantities}\label{Linear polarization degree}
We defined the HEZD model as an optically thin, geometrically narrow ring. Therefore, when we considered only single-scattering events, the net scattered-light polarization was calculated by applying the Stokes formalism analytically. To simulate the radiative transfer in the generally optically thick planetary atmosphere we used the numerical 3D Monte Carlo solver POLARIS (\citealt{Polaris}; \citealt{Lietzow}). We considered scattering of unpolarized stellar radiation by the atmospheric gas as well as spherical cloud and dust particles as polarization mechanism in this study. For a more detailed model description see \Cref{Model}.\\
\\
The observable polarized radiation of the considered system results from the superposition of the direct stellar flux and the thermally reemitted and scattered radiation of the HEZD and the planet. The unpolarized stellar flux dominates the net flux and thus the level of the observable polarization. For practical purposes in the context of our subsequent studies, we thus defined the observable wavelength-dependent linear polarization degree once including the stellar flux (\(P_{\star}\)) and once without (\(P\)):
\begin{align}\label{pdegree}  P_{\star}(\lambda)=\dfrac{F_{\text{pol}}(\lambda)}{F_{\text{sca}}(\lambda)+F_{\text{therm}}(\lambda)+F_{\star}(\lambda)},\end{align}
\begin{align}\label{stardegree} P(\lambda)=\dfrac{F_{\text{pol}}(\lambda)}{F_{\text{sca}}(\lambda)+F_{\text{therm}}(\lambda)}, \end{align}
denoting the flux of the linearly polarized radiation as \(F_{\text{pol}}\), the scattered flux as \(F_{\text{sca}}\), the thermally reemitted flux as \(F_{\text{therm}}\) and the stellar flux as \(F_{\star}\) (see \Cref{Model} for the individual definitions).\\

\subsection{Reference star}\label{Reference star}
Motivated by the presumable detection of polarized radiation in the WASP-18 system from \citet{Bott}, we defined a star of spectral type F6 with radius \(R_{\star} =1.24\ R_{\odot}\) and a stellar effective temperature of \(T_{\star}=6350\) K, with \(d_{\text{obs}}=126.4\ \)pc as distance to the observer, with the same stellar parameters as
the illuminating reference source of the system. While we neglected the possible intrinsic stellar polarization in the case of this reference star in our simulations, we compared the polarization degrees of HEZD and the close-in planet with the measured intrinsic stellar polarization of the Sun and selected FGK dwarfs in the wavelength range \(\lambda\) from 430 nm to 600 nm in \Cref{stellar}.

\subsection{Reference close-in planet}\label{Reference close-in planet}
The properties of our reference close-in planet are motivated by polarimetric measurements of the hot Jupiter WASP-18b for which an upper boundary of the polarization degree (40 ppm) has most likely been detected (\citealt{Bott}). Thus, we chose a planetary radius of \(R_{\text{planet}}=1.16\ R_{\rm{Jup}}\), an orbital radius of \(d_{\text{planet}}=0.02\) au and an equilibrium temperature of \(T_{\text{planet}}=2411\) K (\citealt{Sheppard2017}). This choice is further supported by the fact that the observed polarization degree in the optical wavelength range of hot Jupiters is most likely in the same order of magnitude as that of WASP-18b (\citealt{Bailey2021}).\\
Because currently available data for hot Jupiters are not sufficient to derive a unique model for their atmosphere (e.g., \citealt{Pluriel}), we neglected the high potentially horizontal temperature contrasts, mixtures of species in ionic, atomic, molecular and condensate phases in atmospheres of hot Jupiters (\citealt{Helling}; \citealt{Fortney}) without addressing any nonequilibrium processes that might affect the atmospheric composition and chemistry. The chemical composition of the atmosphere of this reference planet was thus chosen to be as simple as possible to obtain qualitative insights into the polarization characteristics.\\ Because Rayleigh scattering from small cloud particles is anticipated to be the dominant source of polarization for hot Jupiters (e.g., \citealt{Bailey2018,Bott}) we considered an atmosphere consisting of molecular hydrogen (\(\text{H}_{2}\)),  and optically thin forsterite cloud particles (\(\text{Mg\(_{2}\)SiO}_{4}\)) with an effective radius of \(r_{\rm{eff}}=0.05\ \mu \)m and an effective variance of \(v_{\rm{eff}}=0.01\). A more detailed description of the pressure structure and cloud formation is given in \Cref{planetary}. Because the variation in the polarization degree and thus the potentially most decisive quantity that allows a distinction between the individual contributions of the planet and the HEZD to the net polarization degree of the system, is expected to show the strongest variations along its orbit in the case of an edge-on orientation of its orbital plane, we considered a transiting planet (orbital inclination of \(i= 90^\circ\)).\\
\\
The simulated maximum polarization degree (about 14 ppm) of the (reference) planet-star system at a wavelength of 470 nm is reached at a planetary phase angle of \(67^\circ\), which is approximately where the observed polarization peaks as well. This consistency supports the plausability of the chosen properties of the reference model for the close-in planet.

\subsection{Reference hot exozodiacal dust}\label{Reference hot exozodiacal dust}
The strong NIR excess of HEZD that was observed for a wide range of stellar spectral types can be modeled by an optically thin narrow dust ring with a total mass (\({\it M}_{\rm{dust}}\)) in a range of about \(0.03\times 10^{-9}\rm{M}_{\oplus}-10.2\times 10^{-9}\rm{M}_{\oplus}\) consisting of graphite grains with radii (\(\it a\)) between \(0.01\ \mu\rm{m}-1.0\ \mu\)m. The dust is typically located at distances of \(0.02\ \rm{au}-0.4\) au from the central star (\(d_{\rm{dust}}\)) corresponding to dust temperatures (\(T_{\rm{dust}}\)) between \(1000\) K\(-\ 2800\) K \citep{Kirchschlager2017, Kirchschlager2018, Kirchschlager2020}.\\
\\
We defined the reference HEZD model to fit the NIR excess of HD 22484 (10 Tau) with the following appropriate parameter configuration (\citealt{Kirchschlager2018}): The dust grains were assumed to consist of graphite, to be compact spheres with a singles radius of \(\em a=\) \(0.71\ \)\(\mu\)m, and to be  distributed within a ring with a radius of \(d_{\text{dust}}=0.04\) au centered at the central star. The corresponding dust temperature was \(T_{\rm{dust}}=1780\) K, which is close to the sublimation temperature of graphite (2000 K; \citealt{Lamo}), and the HEZD mass was \(\em M\)\(_{\text{dust}}=1.23\times 10^{-9} \text{M}_{\oplus}\). Because the observed NIR excesses of HEZD can also partly be modeled with small dust grains whose temperatures are above this sublimation temperature, the aspect of sublimation was neglected in the context of this study. In addition, the orbital inclination was set  to \(i=90^{\circ}\) (edge-on). Corresponding to the observational constraints we used for the graphite mixture a density of \(\rho = 2.24\ \text{g}\ \text{cm}^{-3}\) (\citealt{Weingartner}) applying the one-third to two-thirds dust grain orientation weighting for the calculation of the optical properties (\citealt{Draine}). \\ 
According to \citet{Absil2013}, a HEZD-star flux ratio of 1.21\% was measured at a wavelength of \(2.13\ \mu\)m for HD 22484. Our defined reference model at the same wavelength agrees by about 87\% with this measured HEZD-star flux ratio, which shows that our model is plausible.
An overview of the parameter values of the reference model is compiled in \Cref{table2}.
\begin{table}[t!]
\caption{Parameters of the reference model consisting of the central star, the close-in planet and the HEZD.}
\label{table2}
\small
\centering
\begin{tabular}{l l c c } 
 \hline 
 & Parameter & Variable & Value \\  
 \hline
 \textbf{Star} & Radius & \(R_{\star}\) &\(1.24\ R_{\odot}\)  \\ 
& Effective temperature & \(T_{\star}\) & 6350 K  \\
& Distance to observer & \(d_{\rm{obs}}\) & 126.4 pc  \\
 \hline
 \textbf{Planet}& Orbital inclination & \(i\) & \(90^\circ\)  \\
& Radius & \(R_{\rm{planet}}\)&\(1.16\ R_{\rm{Jup}}\)\\
& Atmospheric composition && \(\text{H}_{2}\),\ \(\text{Mg\(_{2}\)SiO}_{4}\) clouds\\
  & Orbital radius &\(d_{\rm{planet}}\)& 0.02 au\\
  & Phase angle &\(\alpha\)& \(67^\circ\)\\
   & Equilibrium temperature &\(T_{\rm{planet}}\)& 2411 K\\
   \hline
\textbf{Dust} & Grain radius &\(\em a\)& \(0.71\ \mu\)m\\
    & HEZD mass & \(\em M_{\rm{dust}}\)& \(1.23\times 10^{-9} \rm{M}_{\oplus}\)\\
      & Radius &\(d_{\rm{dust}}\)& 0.04 au\\
     & Temperature &\(T_{\rm{dust}}\)& 1780 K\\  
 \hline
\end{tabular}
\end{table}
\section{Results}\label{results1}
In \Cref{wavelength-dependence} we compare the polarization characteristics between the net polarization degree of the radiation scattered by the close-in exoplanet (hereafter: planetary polarization) and the net polarization degree of the radiation scattered by the HEZD (hereafter: dust polarization) for various representative model configurations. We consider the wavelength-dependent polarization characteristics of a HEZD and a planet for selected parameters to identify characteristic features. We vary the dust grain radius and orbital inclination in \Cref{sizes} and the composition of the atmosphere and planetary phase angle in \Cref{atmosphere}. Moreover, we compare dust and planetary polarization with existing measurements of intrinsic stellar polarization in \Cref{stellar}.\\
In \Cref{V-band} we investigate a system consisting of a close-in planet and a HEZD to study the impact of HEZD on polarimetric investigations of close-in planets for selected wavelengths representative for the photometric bands \(V, K\) and \(L\). We vary the following parameters for a comparison of the resulting linear polarized fluxes: The dust grain radius and the HEZD mass (\Cref{masses}) and the composition of the planetary atmosphere and phase angle (\Cref{phase}). We also investigate the potential of analyzing the individual Stokes parameters \(Q\) and \(U\)  (\Cref{Q_U}). Unless stated otherwise, we applied the parameter setup of the reference model (see \Cref{table2}). The discussion is conducted under the assumption of ideal observations without consideration of possible technical limitations.

\subsection{Wavelength-dependent polarization}\label{wavelength-dependence}
The following discussion focuses on individual features of the wavelength-dependent polarization for a HEZD and a planet in the wavelength range from 550 nm to \(4\ \mu\)m. To derive conclusions about the polarization characteristics independent of the HEZD mass, the radius of the HEZD and planetary orbital radius, we consider the polarization as a function of the wavelength relative to the polarization at a wavelength of 550 nm that is representative for the photometric \(V\) band, in the vicinity of which previous polarimetric observations of close-in exoplanets or HEZD have been performed (e.g., \citealt{Marshall2016}; \citealt{Bott}).
\subsubsection{Characterization of wavelength-dependent polarization}\label{referenceplot}
\begin{figure}[t!]
\includegraphics[scale=0.20]{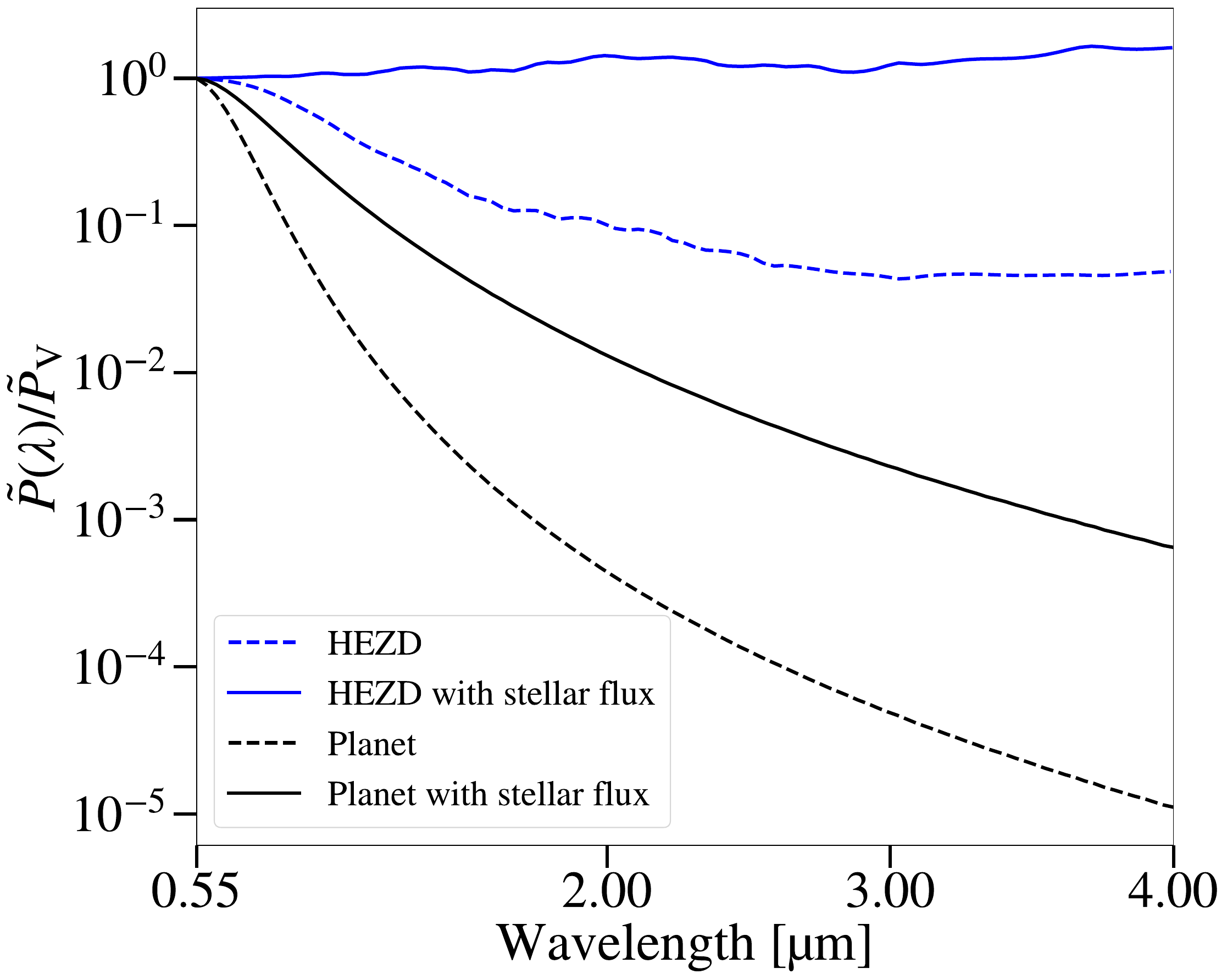}
\caption{Polarization as a function of wavelength relative to the polarization at a wavelength of 550 nm for a HEZD and a planet (reference model; \Cref{table2}) with (\(\tilde{P}(\lambda)=P_{\star}(\lambda)\), \(\tilde{P}_{V}=P_{\star V}\)) and without (\(\tilde{P}(\lambda)=P(\lambda)\), \(\tilde{P}_{V}=P_{V}\)) the stellar flux to the net polarization degree. See \Cref{referenceplot} for details.}\label{Fig 1}
\end{figure}

\begin{figure}[t!]
\includegraphics[scale=0.20]{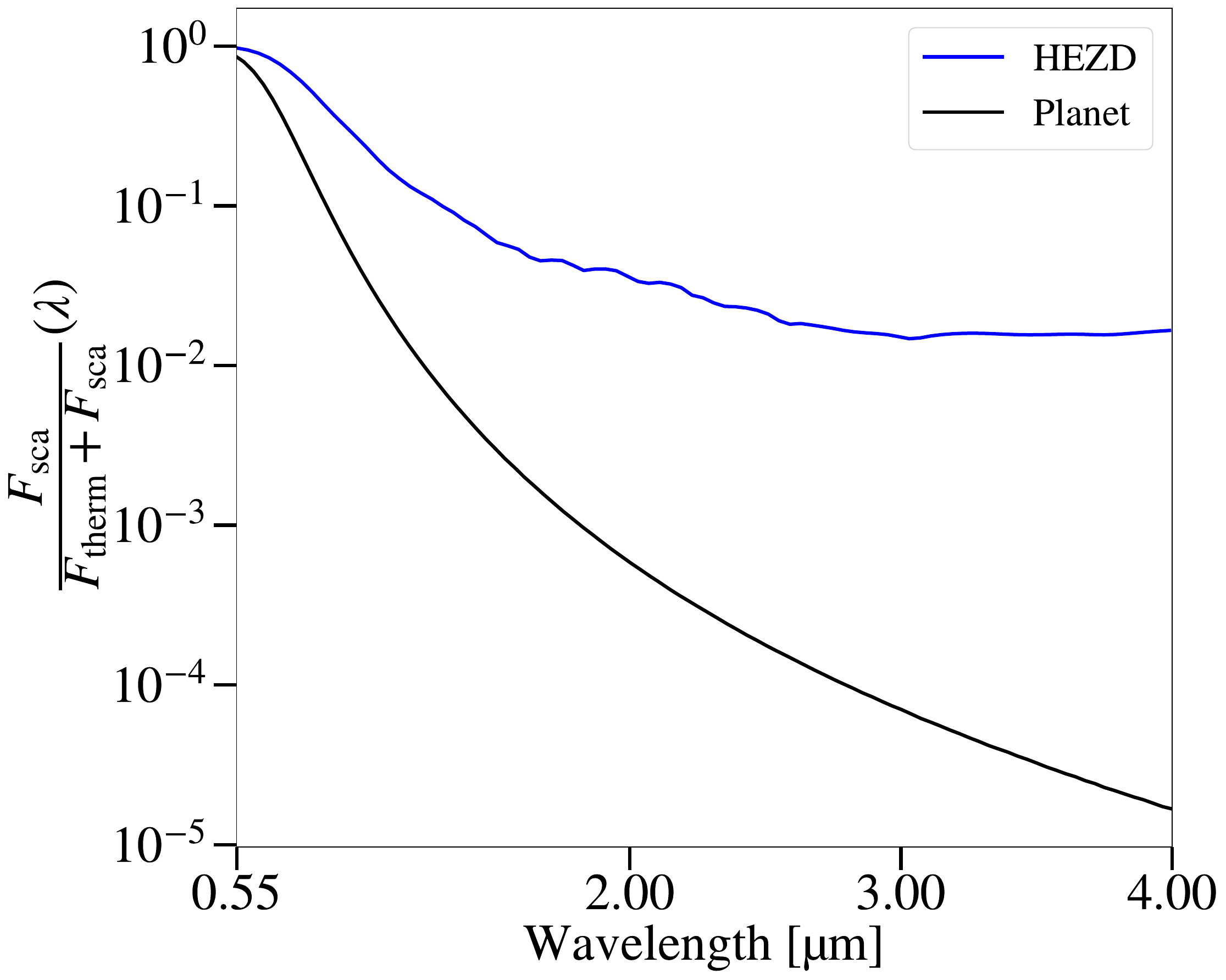}
\caption{Relative contribution of the scattered flux of a HEZD and a planet to the net flux (scattered and thermally reemitted flux) of both components (reference model; \Cref{table2}). See \Cref{referenceplot} for details.} \label{Fig 2}
\end{figure}

In the case of the planet the polarization ratio \(P(\lambda)/P_{V}\) decreases steeply within the considered wavelength range (see \Cref{Fig 1}). In the case of the HEZD the ratio decreases with minor fluctuations, but less steeply than that of the planet. The thermal flux \(F_{\text{therm}}\) of the planet and the HEZD is higher in the NIR wavelength range than the scattered flux \(F_{\text{sca}}\) (see \Cref{Fig 2}), resulting in lower polarization ratios in the NIR wavelength range. The ratio of the planetary scattered flux to the total planetary flux is similar to that in observations from \citet{Krenn} for low geometric albedos of hot Jupiters.\\ The wavelength-dependent polarization characteristics result from the optical properties of dust grains and atmospheric particles. In the planetary atmosphere, Rayleigh scattering by \(\text{H}_{2}\) molecules and small \(\text{Mg\(_{2}\)SiO}_{4}\) cloud particles determines the wavelength-dependent polarization, while it is Mie scattering in the case of the HEZD (for a detailed discussion of the influence of the dust grain radius on the polarization characteristics see \Cref{sizes}). When the direct unpolarized stellar radiation is taken into account, the resulting polarization ratio \(P_{\star}(\lambda)/P_{\star V}\) is affected as well (see \Cref{Fig 1}). As the stellar flux has its maximum in the optical wavelength range close to the \(V\) band, the polarization ratios show a flatter decline in the NIR wavelength range. The impact of central stars with spectral types A and G on the net wavelength-dependent polarization characteristics of the star-planet-HEZD system is given in \Cref{Spectral}. We conclude that a distinction between a planet and a HEZD based on their wavelength-dependent polarization characteristics for the considered parameters is thus possible by comparing the ratios of polarization degree measurements in the optical and NIR wavelength range.

\subsubsection{Dust grain radius and inclination}\label{sizes}
\begin{figure}[t!]
\begin{minipage}{0.54\textwidth}
\includegraphics[width=\textwidth]{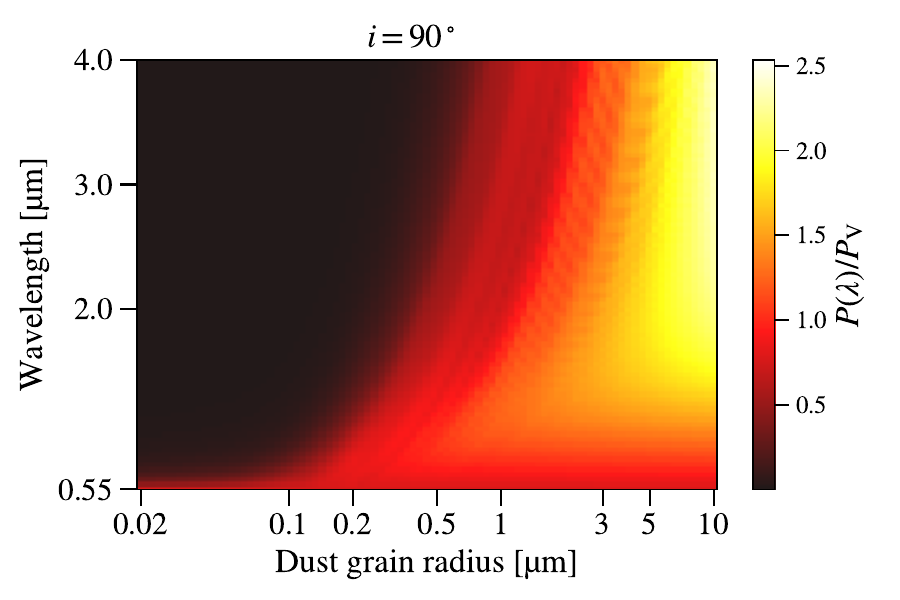}
\end{minipage}
\begin{minipage}{0.54\textwidth}
\includegraphics[width=\textwidth]{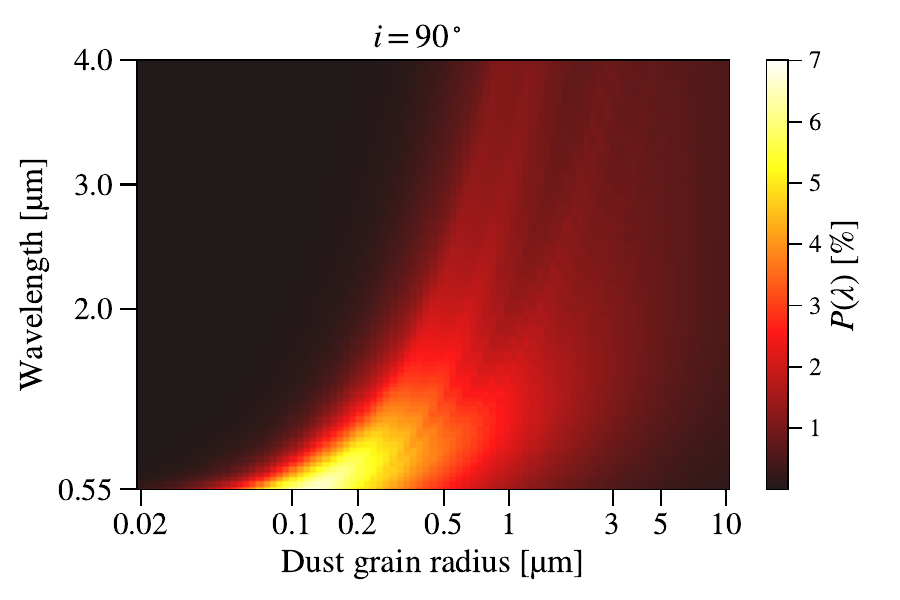}

\end{minipage}
\caption{Dust polarization for different wavelengths and dust grain radii. Top: Dust polarization as a function of wavelength and dust grain radius for an orbital inclination of \(90^\circ\). Bottom: Dust polarization as a function of wavelength and dust grain radius relative to the polarization at a wavelength of 550 nm for an orbital inclination of \(90^\circ\). See \Cref{sizes} for details.}\label{Fig 3}
\end{figure}  

\begin{figure}[t!]
\includegraphics[scale=0.20]{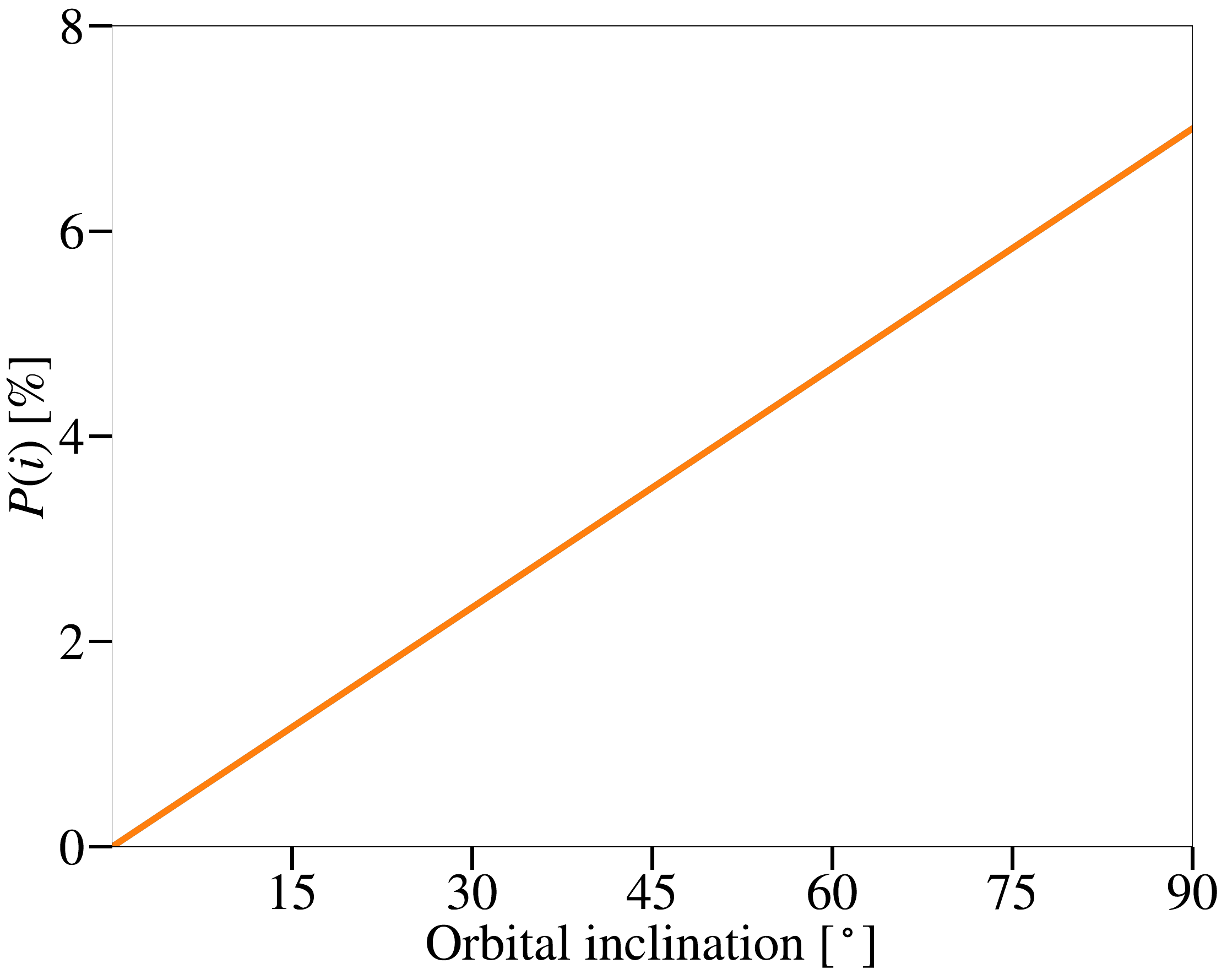}
\caption{Dust polarization as a function of orbital inclination at a wavelength of 550 nm for a dust grain radius of \(0.1\ \mu\)m. See \Cref{sizes} for details.} \label{Fig 4}
\end{figure}
 
Although the wavelength dependence of the NIR excess of a HEZD is best explained by the thermal reemission of submicrometer-sized dust grains \citep{Kirchschlager2017,Kirchschlager2018,Kirchschlager2020}, the lack of detections and stringent constraints at longer wavelengths in the MIR to millimeter range, does not allow an exclusion of larger particles so far (\citealt{Stuber}). For this reason, particle radii up to \(10\ \mu\)m are also considered in the following discussion of the impact of the dust grain radius for a fixed HEZD mass on the wavelength-dependent polarization characteristics. The impact of a dust grain size distribution on the wavelength-dependent polarization characteristics is given in \Cref{distribution}. We identify the following three trends (see \Cref{Fig 3}):\\
First, for dust grain radii smaller than  \(0.1\ \mu\)m and wavelengths longer than 550 nm the polarization drops to below \(1\%\) (top plot of \Cref{Fig 3}) because small dust grains have a higher temperature than larger dust grains at the same orbital distance from the central star. Their thermal reemission therefore dominates the net flux in the NIR wavelength range, which strongly reduces the resulting net polarization. Consequently, the contribution of scattered radiation is negligible for these small grains. Second, the highest polarization is about \(7\%\) at a wavelength of 550 nm for a dust grain radius of \(0.1\ \mu\)m, which has a strong impact on the polarimetric analysis of close-in planets in the presence of a HEZD (see \Cref{V-band}). Third, at the maximum considered wavelength and dust grain radius (\(\lambda=4\ \mu\)m, \(\em a=10\ \mu\)m), the polarization ratio reaches its highest value of about 2.5 (bottom plot of \Cref{Fig 3}) because the scattering efficiency increases at infrared wavelengths (larger dust grains are redder in their scattering color). We furthermore note that the dust polarization degree in the \(K\) band agrees well with the findings of \citet{Kirchschlager2017}.\\ In \Cref{Fig 4} the dependence of the dust polarization on the orbital inclination of the HEZD ring is shown for dust grains with a radius of \(0.1\ \mu\)m at a wavelength of 550 nm. These particular values of the grain radius and wavelength were chosen because the dust polarization was found to reach its maximum value in this case. The net polarization increases with increasing inclination and reaches its maximum of about \(7\%\) at \(90^\circ\). Because of symmetry constraints, the net polarization equals zero in the case of zero inclination (face-on).\\ In summary we find that while the orbital inclination affects the net dust polarization, the dust grain radius has a strong impact on the wavelength-dependent polarization characteristics and also on the net polarization. It therefore strongly affects the polarimetric investigation of a close-in planet in the presence of a HEZD. For small dust grain radii the resulting net polarization of a HEZD is negligible within the wavelength range from 550 nm to \(4\ \mu\)m for a polarimetric analysis like this. Furthermore, polarimetric measurements of HEZD in the NIR wavelength range and in the \(V\) band potentially provide constraints on the dust grain radius by a comparison of the polarization ratios.
\newpage
\subsubsection{Composition of the planetary atmosphere}\label{atmosphere}
\begin{figure}[t!]
\begin{minipage}{0.47\textwidth}
\includegraphics[width=\textwidth]{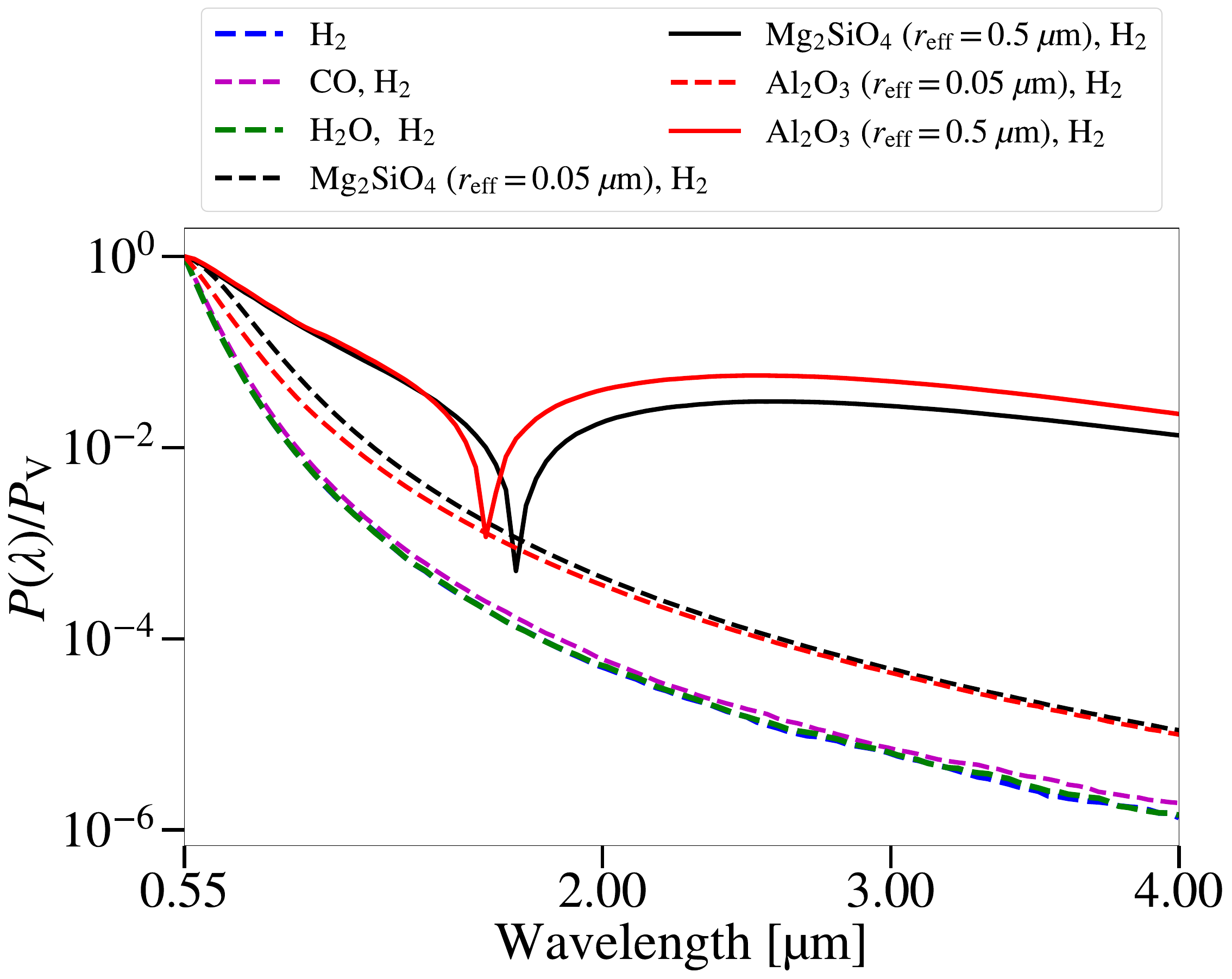}

\end{minipage}
\begin{minipage}{0.47\textwidth}
\includegraphics[width=\textwidth]{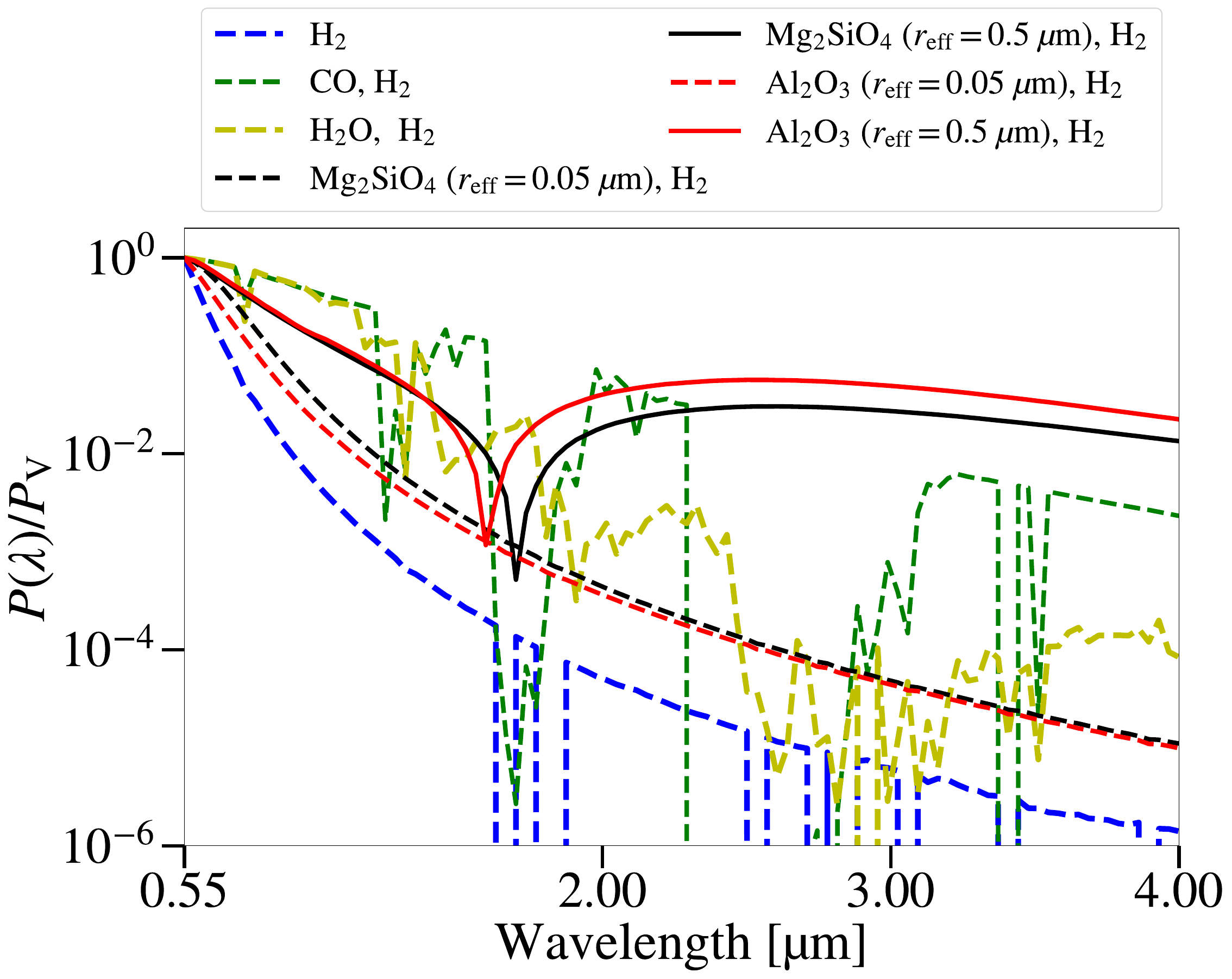}

\end{minipage}
\caption{Planetary polarization as a function of wavelength relative to the polarization at a wavelength of 550 nm for compositions of the planetary atmosphere (\(\text{H}_{2}\), \(\text{CO}\), \(\text{H}_{2}\text{O}\), \(\text{Mg\(_{2}\)SiO}_{4}\) and \(\text{Al}_{2}\text{O}_{3}\) clouds). Bottom and top: With and without the atmospheric specific absorption coefficients. See \Cref{atmosphere} for details.}\label{Fig 5}
\end{figure}  
\begin{figure}[t!]
\includegraphics[scale=0.21]{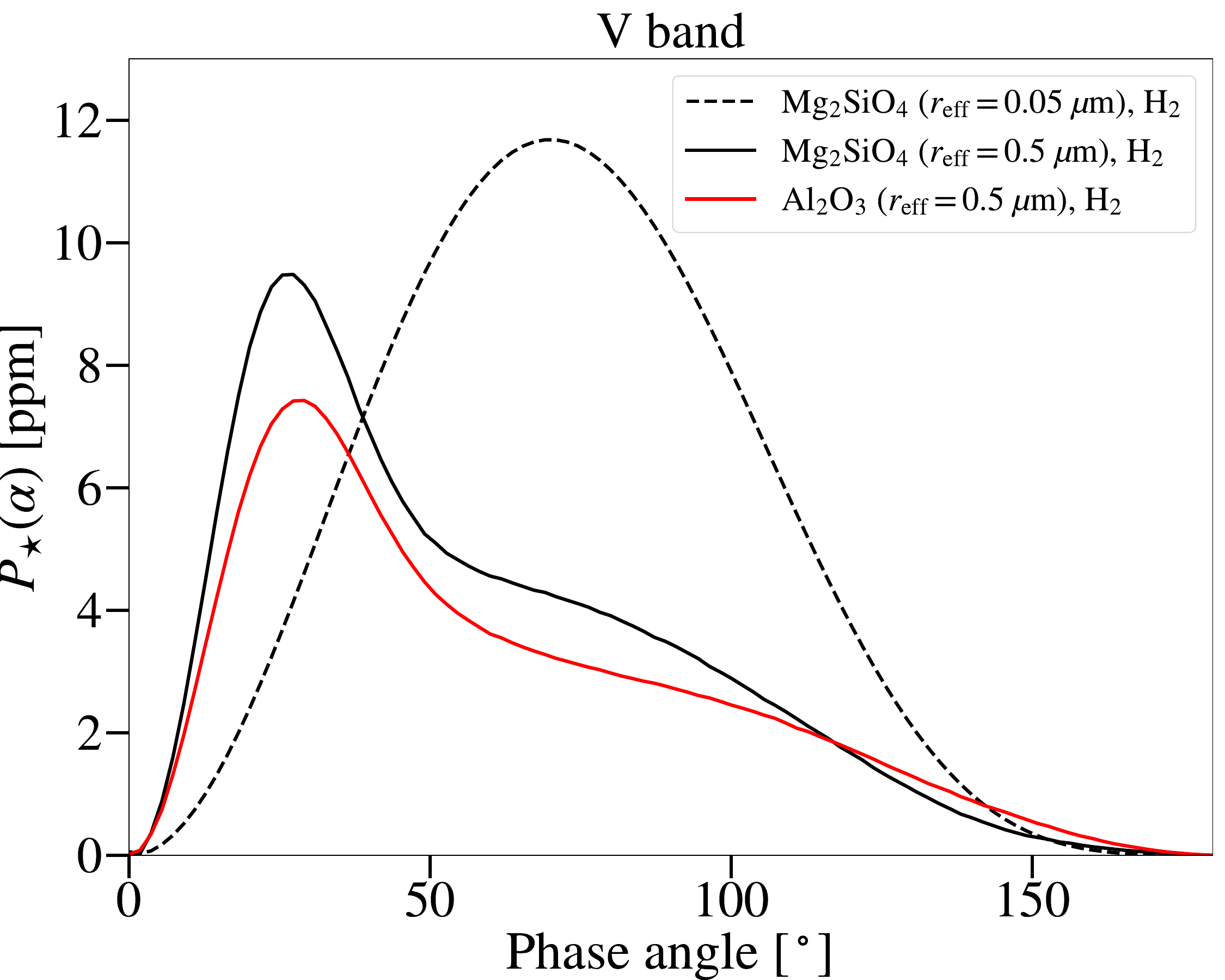}
\caption{Planetary polarization as a function of phase angle for compositions of the planetary atmosphere (\(\text{H}_{2}\), \(\text{Mg\(_{2}\)SiO}_{4}\) and \(\text{Al}_{2}\text{O}_{3}\) clouds) at a wavelength of 550 nm. See \Cref{atmosphere} for details.}\label{Fig 6}
\end{figure}

We illustrate the planetary polarization as a function of wavelength relative to the polarization in the \(V\) band for selected atmospheric compositions to investigate their impact on the planetary polarization characteristics (see \Cref{Fig 5}). The considered mixing ratios (\(\text{CO}/\text{H}_{2}=0.2\), \(\text{H}_{2}\text{O}/\text{H}_{2}=10^{-3})\) were adopted from \citet{Sheppard2017} to represent the atmospheric compositions of WASP-18b and the effective radii \(r_{\rm{eff}}=0.05\ \mu \)m, \(0.5\ \mu \)m and effective variances \(\nu_{\rm{eff}}=0.01, 0.1\ \) for \(\text{Mg\(_{2}\)SiO}_{4}\) and \(\text{Al}_{2}\text{O}_{3}\) cloud particles are adopted from \citet{Bailey2018}, \citet{Bott} and \citet{Kedziora} (see \Cref{planetary} for the size distribution of the cloud particles). Although WASP-18b differs from many other hot Jupiters in terms of its temperature and planetary orbital radius, the results presented here can be generalized for hot Jupiters because the selected atmospheric compositions, resulting in Rayleigh or Mie scattering, also apply to other extensively studied hot Jupiters such as HD 189733b (\citealt{Sing2015}; \citealt{Wakeford2015}; \citealt{Mansfield}) and are thus expected to result in at least qualitatively similar polarization characteristics.\\ Three scattering scenarios contribute in this case: Rayleigh scattering on atmospheric particles, Rayleigh and Mie scattering on cloud particles. When no absorption of stellar radiation for the atmospheric particles was considered (see top graph of \Cref{Fig 5}) the polarization ratios of the  \(\text{CO}/ \text{H}_{2}\), \(\text{H}_{2}\text{O}/ \text{H}_{2}\) and \(\text{H}_{2}\) compositions were close to the ratio of the reference atmospheric composition (\(\text{H}_{2}\) atmosphere and \(\text{Mg\(_{2}\)SiO}_{4}\) cloud particles with an effective radius of \(0.05\ \mu \)m) and thus have similar wavelength-dependent polarization characteristics. The reason is that Rayleigh scattering has the strongest impact on the resulting polarization characteristics for the selected compositions because the radii of the gas molecules are small. Moreover, scattering on \(\text{H}_{2}\) is the source of polarization in the clear atmosphere case and the dominant gas for the other mixing ratios. Rayleigh scattering also applies to the \(\text{Mg\(_{2}\)SiO}_{4}\) and \(\text{Al}_{2}\text{O}_{3}\) clouds with an effective radius of \(0.05\ \mu \)m and an effective variance of \(0.01\).\\ The polarization in the NIR wavelength range is significantly lower than in the \(V\) band for every selected atmospheric composition, especially for the Rayleigh-scattering particles. Additionally, the polarization increases in the NIR wavelength range for cloud particles with an effective radius of \(0.5\ \mu \)m and an effective variance of \(0.1\) because Mie scattering applies for particles of such radii. Due to the strong similarities of the polarization ratio for the selected atmospheric compositions, drawing unambiguous conclusions about the composition of the atmosphere only from the measured polarization is hardly feasible, at least as long as small cloud particles dominate the composition of the atmosphere.\\
\\
In addition, we also considered the specific molecular absorption of stellar radiation of the atmospheric gas particles (see bottom graph of \Cref{Fig 5}). These absorption features are imprinted on the wavelength-dependent polarization, resulting in significant differences between the polarization ratios for the clear atmospheres with different chemical compositions, potentially in the NIR wavelength range. However, for atmospheres with clouds these features are strongly damped because the absorption of the gas particles is negligible in comparison to the absorption by the cloud particles within the model we considered.\\ Motivated by the wavelength-dependence of the planetary polarization for different effective cloud particle radii, we illustrate the polarization degree for three different atmospheric compositions as a function of phase angle in \Cref{Fig 6}. When the effective cloud particle radii are increased to \(0.5\ \mu\)m, the \(\text{Mg\(_{2}\)SiO}_{4}\) and \(\text{Al}_{2}\text{O}_{3}\) particles move into the Mie scattering regime, which results in a substantially reduced polarization degree and a different phase angle dependence of the planetary polarizaton in comparison to the Rayleigh scattering regime (maximum polarization is now at about \(40^\circ\)) because the peak of the back-scattering probability is narrower.\\
In conclusion, a comparison of the characteristic wavelength-dependent polarization ratios for dust and atmospheric particles enables a distinction of the polarization characteristics for the HEZD and the planet, whereby the resulting polarization of large cloud particles potentially mimic the wavelength-dependent polarization characteristics of large dust grains (Mie scattering) and likewise with respect to small dust grains and atmospheric particles (Rayleigh scattering). Within the considered parameter space, an additional consideration of the molecule-specific absorption coefficients for the wavelength-dependent polarization characteristics enables a unique characterization of the atmospheric and dust particles in the case of clear atmospheres.

\subsubsection{Intrinsic stellar polarization}\label{stellar}
\begin{figure}[t!]
\includegraphics[scale=0.20]{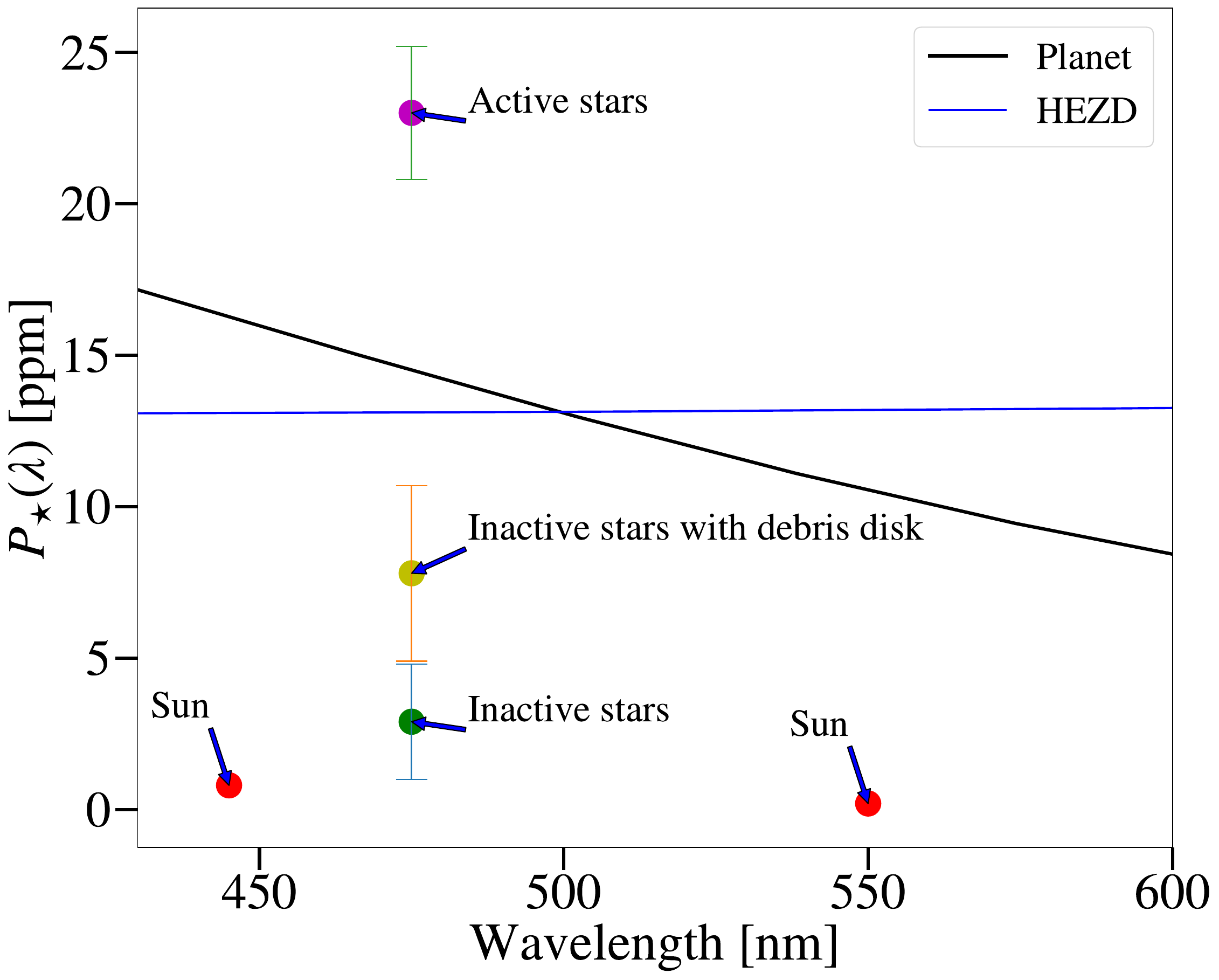}
\caption{Polarization as a function of wavelength for a HEZD and a planet (reference model; \Cref{table2}). Two measured polarization degrees of the Sun (red dots; \citealt{Kemp}) and average polarization degrees for three types of observed FGK dwarfs are also shown (green, yellow and purple dots; \citealt{Cotton}). See \Cref{stellar} for details.} \label{Fig 7}
\end{figure}
To investigate whether the intrinsic stellar polarization affects the polarimetric analysis of a HEZD and a planet (see \Cref{table2} for the model parameters), we compared their wavelength-dependent polarization degrees in the range from 430 nm to 600 nm with polarization measurements of the Sun and of selected FGK dwarfs (see \Cref{Fig 7}). The maximum dust polarization for parameters from the reference model is about 13 ppm while the maximum planetary polarization is about 17 ppm (both at a wavelength of 430 nm).\\ The average polarization degree for a sample of nearby FGK dwarfs (\citealt{Cotton}) was divided into three classes: The first class consists of inactive stars with an average polarization degree of about 2.9 ppm. The second class consists of inactive stars with debris disks with an average polarization degree of about 7.8 ppm. Here, the additional polarization of about 4.9 ppm in comparison to the inactive stars without debris disk results from scattering of stellar radiation by the debris disk. The third class contains active stars with an average polarization degree of about 23 ppm exceeding even the maximum planetary polarization. These high intrinsic stellar polarization degrees presumably result from differential saturation that is also induced by the global magnetic field, whereby differential saturation means that many spectral lines overlap and merge with each other to produce a net broadband linear polarization \citep{Cotton, Cotton2019}. Measured polarization degrees of the Sun are also shown in Fig. 7 (\citealt{Kemp}) at about 0.8 ppm at 450 nm and about 0.2 ppm at 550 nm, which fits into the order of magnitude of averaged polarization for inactive stars without a debris disk (first class of sampled FGK dwarfs). In summary, the stellar polarization contributes significantly to the total polarization of the system, especially for active stars. In conclusion, an analysis of the scattered-light polarization for close-in planets orbiting active stars in the presence of a HEZD is therefore challenging and requires independent constraints on the intrinsic wavelength-dependent stellar polarization. Time domain polarimetry represents a promising approach to disentangle stellar and planetary signals based on their potentially different temporal dependence, similar to the established technique of high-precision long-duration photometry (e.g., \citealt{Bruno}).
\subsection{Contribution of hot exozodiacal dust to the total polarized flux}\label{V-band}
We studied to which extent the net flux of the radiation polarized by the HEZD (hereafter: polarized flux of the dust \(F_{\mathrm{pol\ dust}}\)) contributes to the net flux of the radiation polarized by the planet (hereafter: planetary polarized flux \(F_{\mathrm{pol\ planet}}\)) at 550 nm, \( 2.22\ \mu\)m, and \(3.45\ \mu\)m, which is representative for the photometric bands \(V, K\) and \(L\). We varied the following parameters: The HEZD mass, the dust grain radius for a fixed HEZD mass (\Cref{masses}), the planetary phase angle and the  composition of the planetary atmosphere (\Cref{phase}).\\
\begin{figure}[t!]
\includegraphics[scale=0.20]{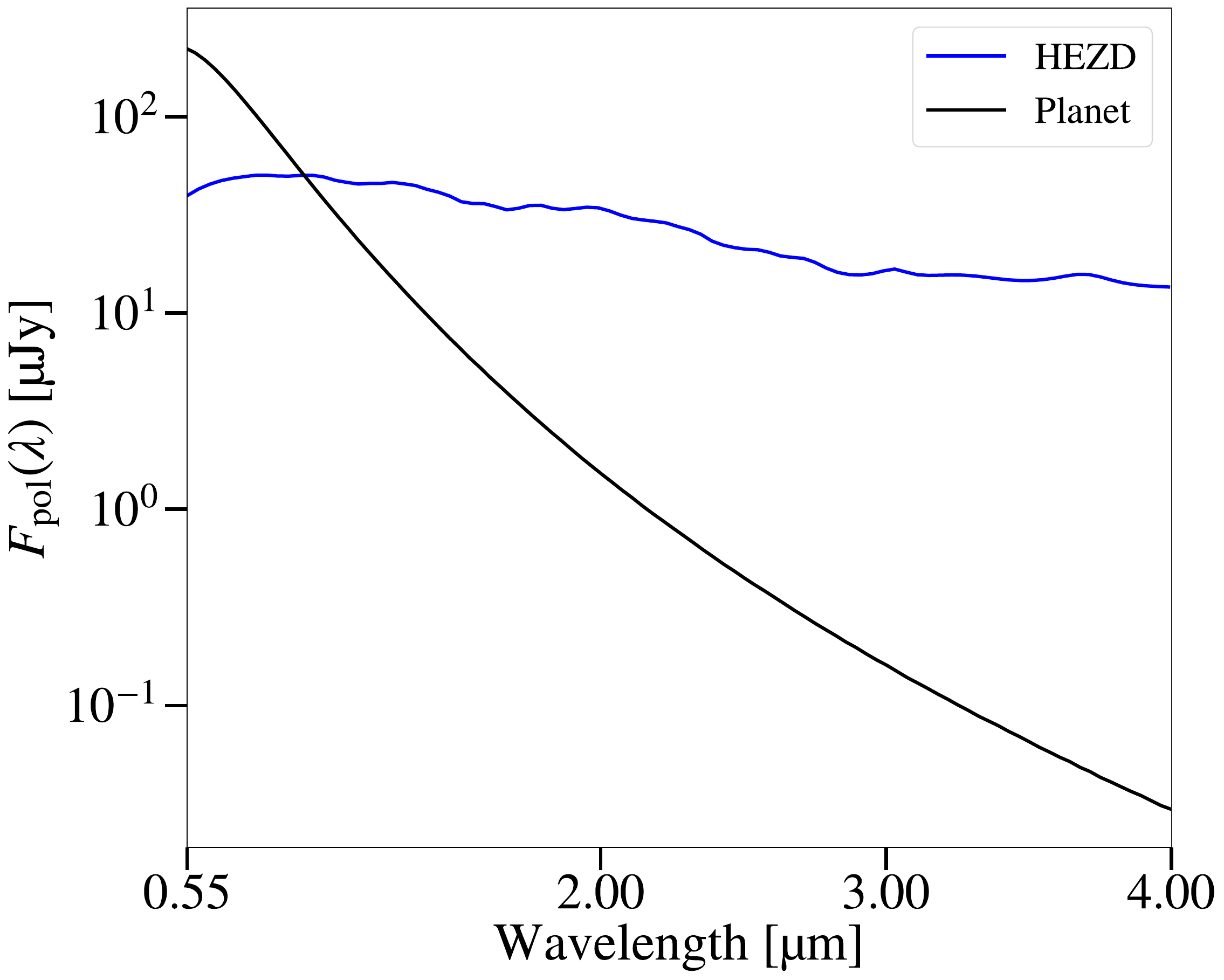}
\caption{Polarized flux as a function of wavelength for a HEZD and a planet (reference model; \Cref{table2}). See \Cref{V-band} for details.} \label{Fig 8}
\end{figure}
\\
For the parameters from the reference model (\Cref{table2}), the maximum planetary polarized flux (about 122 \(\mu\)Jy) is higher than the maximum polarized flux of dust (about 39 \(\mu\)Jy), but it decreases much faster (see \Cref{Fig 8}). This is due to Mie scattering in the case of the HEZD and Rayleigh scattering in the case of the planetary atmosphere. For polarimetric observations in the NIR wavelength range, the influence of the planet is therefore negligible within the considered parameter range.

\subsubsection{Hot exozodiacal dust mass and dust grain radius}\label{masses}
\begin{figure}[t!]
\begin{minipage}{0.51\textwidth}
\includegraphics[width=\textwidth]{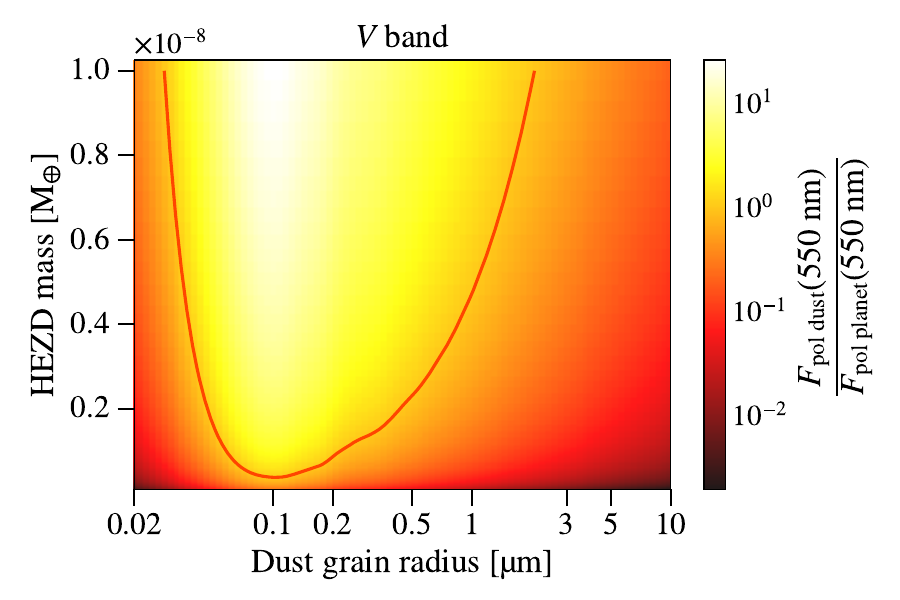}

\end{minipage}
\begin{minipage}{0.51\textwidth}
\includegraphics[width=\textwidth]{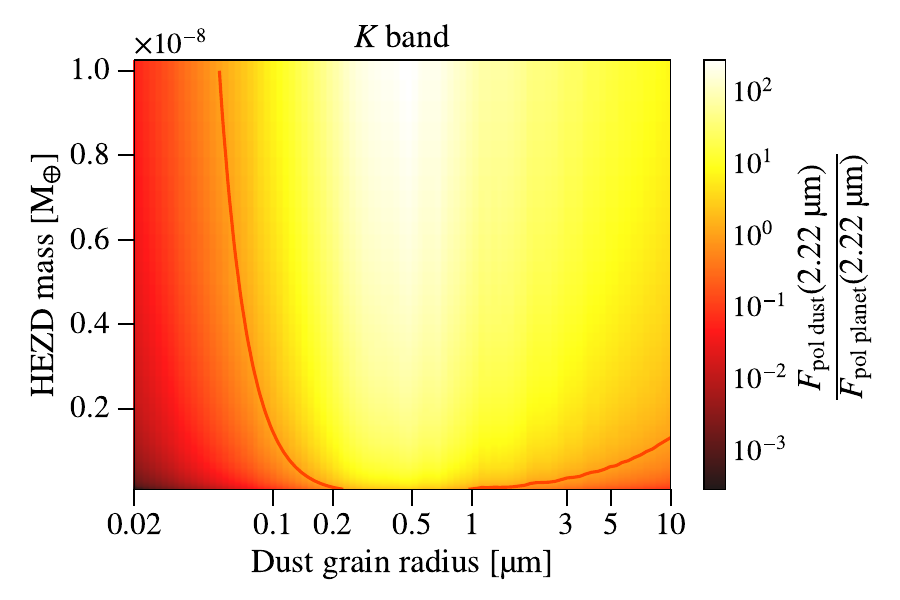}

\end{minipage}
\begin{minipage}{0.51\textwidth}
\includegraphics[width=\textwidth]{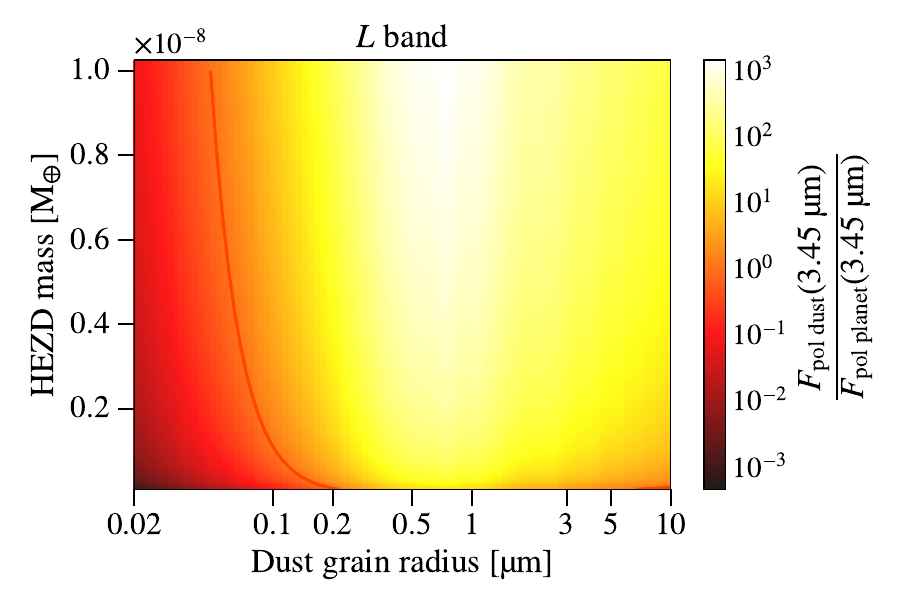}

\end{minipage}
\caption{Polarized flux of the dust as a function of HEZD mass and dust grain radius relative to the planetary polarized flux (reference model; \Cref{table2}). These relations are shown at a wavelength of 550 nm (top), \(2.22\ \mu\)m (middle) and \(3.45\ \mu\)m (bottom). See \Cref{masses} for details.} \label{Fig 9}
\end{figure}

Because current constraints on the HEZD masses cover a wide range (around \(10^{-10}\)\(\rm{M}_{\oplus}\)  \(-\ 10^{-8}\)\(\rm{M}_{\oplus}\)) and the dust grain radius has a strong impact on the wavelength-dependent polarization characteristics (see \Cref{sizes}), we studied the effect of combining these two parameters on the polarized flux of the dust. Furthermore, we investigated the contribution of the polarized flux of the dust relative to the planetary polarized flux. We find three trends (see \Cref{Fig 9}):\\
At first, for the selected HEZD model, the required dust grain radius to reach the maximum polarized flux of the dust increases from 0.1 \(\mu\)m at a wavelength of 550 nm, over 0.5 \(\mu\)m at a wavelength of \(2.22\ \mu\)m, to 0.8 \(\mu\)m at a wavelength of \(3.45\ \mu\)m. Second, the ratio of the polarized flux of the dust and planetary polarized flux increases from a maximum ratio of about 10 at a wavelength of 550 nm, over about 100 at \(2.22\ \mu\)m, to about 1000 at \(3.45\ \mu\)m because the planetary polarized flux decreases faster than the polarized flux of the dust with increasing wavelength for the selected model parameters.\\ Third, the minimum HEZD mass required for the polarized flux of the dust to exceed the planetary polarized flux (\(F_{\mathrm{pol\ dust}}/F_{\mathrm{pol\ planet}}>1\)) for a dust grain radius of 0.1 \(\mu\)m (see the highlighted orange contour line in  \Cref{Fig 9}) is about \(0.08\times 10^{-8} \rm{M}_{\oplus}\) at a wavelength of 550 nm. The minimum HEZD mass required for the polarized flux of the dust to exceed the planetary polarized flux at  \(2.22\ \mu\)m and \(3.45\ \mu\)m  for certain dust grain radii is even below the lowest considered HEZD mass of \(10^{-10}\ \rm{M}_{\oplus}\). To increase the polarized flux of the dust for a fixed dust grain radius, a higher HEZD mass and thus more particles of the same dust grain radius are required.\\
\\
The orbital radius of a HEZD as well as the planetary orbital radius cover wide ranges (for \(d_{\rm{dust}}\) around \(0.02\ \rm{au}-0.4\) au and for \(d_{\rm{planet}}\) around \(0.01\ \rm{au}-0.05\) au). For both fixed parameters that characterize the planet and the HEZD, their relative individual contributions are almost exclusively determined by the geometrical dilution of the stellar radiation before scattering (the minor impact due to different illumination geometries at different distances from the central star is neglected at this point). We express the polarized flux of the dust relative to the polarized planetary flux as a function of these parameters (\(d_{\rm{dust}}\), \(d_{\rm{planet}}\)) with all other model parameters fixed. We define this relation for the dust grain radii resulting in the highest polarized flux of the dust at the wavelengths 550 nm, \(2.22\ \mu\)m and \(3.45\ \mu\)m and for other model parameters selected according to the reference model (see \Cref{table2}) as
\begin{equation} \label{relation}  \frac{F_{\rm{pol\ dust}}\left(\lambda, d_{\rm{dust}}\right)}{F_{\rm{pol\ planet}}(\lambda,d_{\rm{planet}})}=j\left(\lambda\right)\cdot \frac{d_{\rm{planet}}^{2}}{d_{\rm{dust}}^{2}}, \end{equation}
with \(j\left(550\ \rm{nm}\right)=0.68,\ j\left(2.22\ \mu\rm{m}\right)=34.28\) and \( j\left(3.45\ \mu\rm{m}\right)=360.\)
In conclusion, the impact of the presence of a HEZD on the polarimetric analysis of a close-in planet increases with increasing HEZD mass, a certain dust grain radius depending on the observed wavelength, increasing radius of the planetary orbit and decreasing orbital radius of a HEZD. Due to the large possible parameter space of a HEZD, one suitable parameter configuration to fit the observed NIR excess in certain cases contains dust particles with radii smaller than \(0.1\ \mu\)m resulting in low polarization. The dust therefore has to be located very close to the star and the HEZD mass has to be sufficiently high in these cases to significantly contribute the total polarization of the system in the case of a close-in planet being present. 

\subsubsection{Phase angle and planetary atmosphere}\label{phase}
\begin{figure}[t!]
\begin{minipage}{0.42\textwidth}
\includegraphics[width=\textwidth]{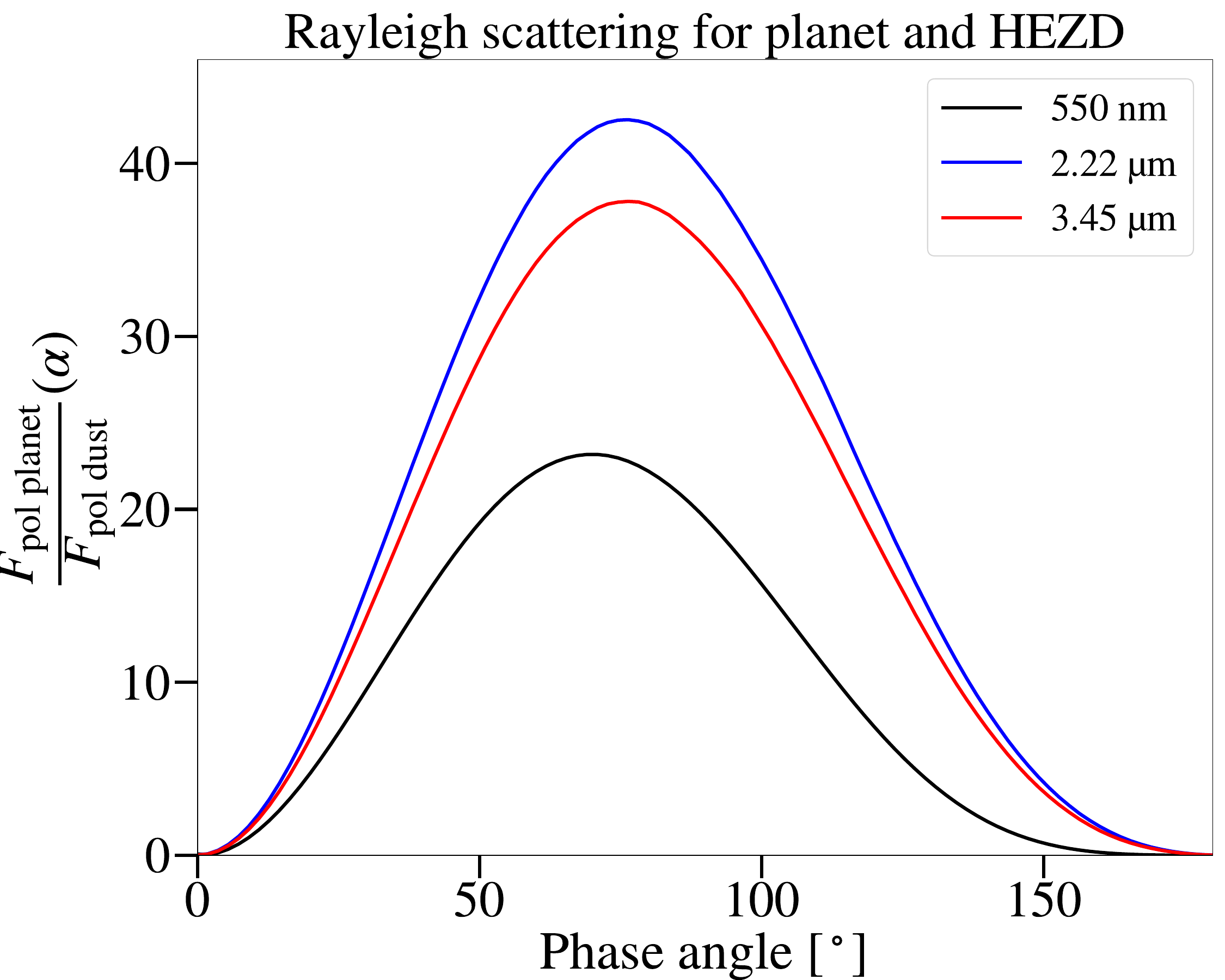}

\end{minipage}
\begin{minipage}{0.42\textwidth}
\includegraphics[width=\textwidth]{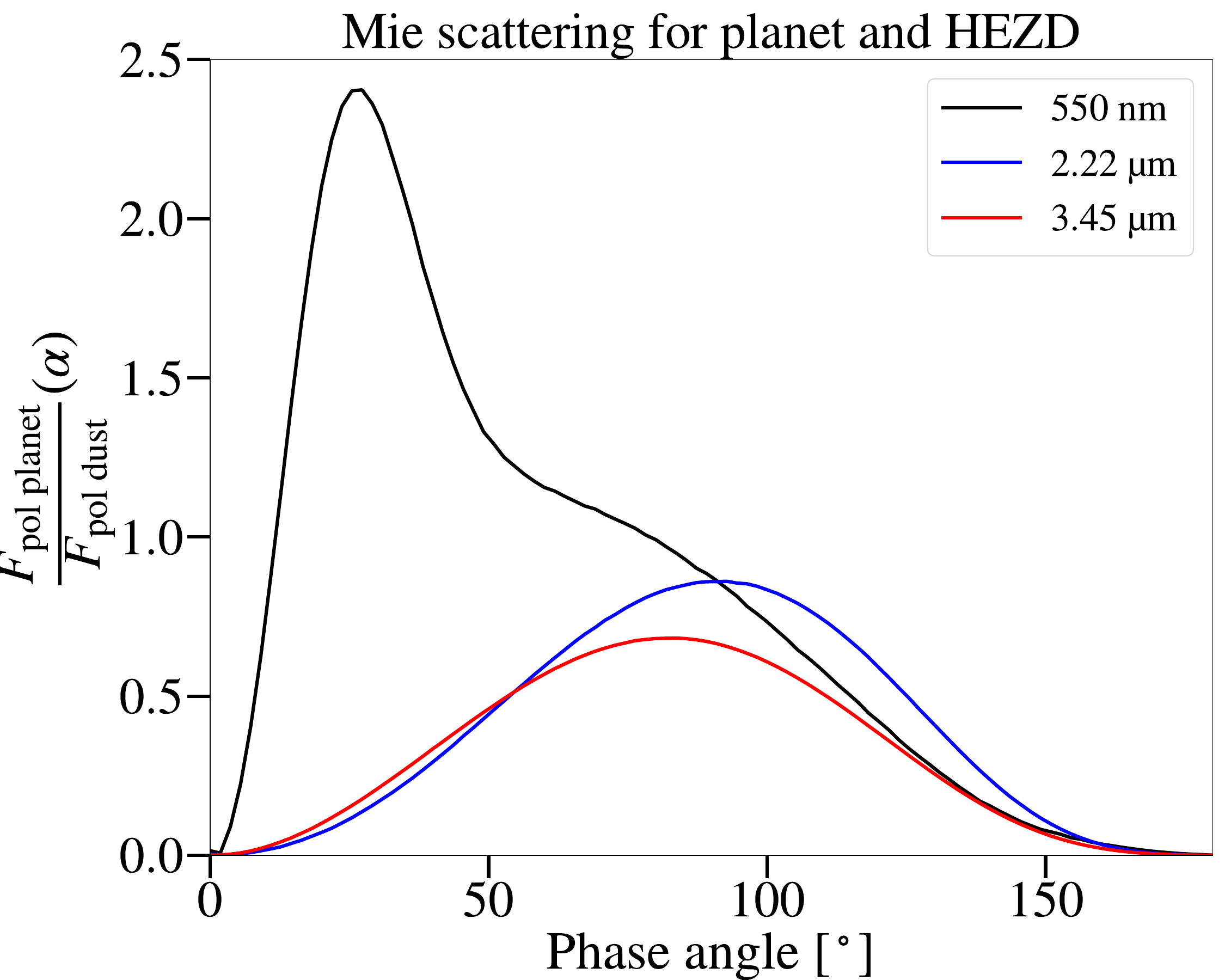}

\end{minipage}
\caption{Planetary polarized flux as a function of phase angle relative to the polarized flux of the dust at the wavelengths 550 nm, \(2.22\ \mu\)m, and \(3.45\ \mu\)m. Top: Planetary atmosphere consisting of \(\text{H}_{2}\) and also \(\text{Mg\(_{2}\)SiO}_{4}\) cloud particles with an effective radius of \(r_{\rm{eff}}=0.05\ \mu \)m. The HEZD consists of dust grains with a radius of \(0.02\ \mu\)m. Bottom: Same chemical composition but for an effective cloud particle radius of \(r_{\rm{eff}}=0.5\ \mu \)m and the HEZD consists of dust grains with a radius of \(0.71\ \mu\)m. All other parameters for the planet and the HEZD are those of the reference model (\Cref{table2}). See \Cref{phase} for details.}\label{Fig 10}
\end{figure}  

Motivated by the findings in \Cref{sizes} and \Cref{atmosphere}, we investigated the influence of the phase angle on the relation between planetary polarized flux and polarized flux of the dust in one example case of Rayleigh scattering (dust grain radius of \(0.02\ \mu \)m and forsterite cloud particles with an effective radius of \(0.05\ \mu \)m) and one example case of Mie scattering (dust grain radius of \(0.71\ \mu \)m and forsterite cloud particles with an effective radius of \(0.5\ \mu \)m) for three selected wavelengths (550 nm, \(2.22\ \mu \)m and  \(3.45\ \mu \)m). We find three interesting trends:\\ First, while in the case of Rayleigh scattering (top graph of \Cref{Fig 10}) the phase angle for the maximum flux ratio deviates only slightly for increasing wavelengths from  \(67^\circ\), the maximum flux ratio shifts to almost \(40^\circ\) in the case of Mie scattering (bottom graph of \Cref{Fig 10}). Second, in the Mie scattering case, the planetary polarized flux only exceeds the polarized flux of the dust (\(F_{\mathrm{pol\ planet}}/F_{\mathrm{pol\ dust}}>1\)) at a wavelength of 550 nm by a factor of about 2.4. However, in the case of Rayleigh scattering, the flux ratio exceeds a value of 10 in the NIR wavelength range and even 40 at 550 nm. Third, the polarized flux of the dust never exceeds the planetary polarized flux in the Rayleigh scattering case for the selected parameters.\\ The atmospheric composition, as well as the phase angle, thus have a significant influence on a polarimetric analysis of a HEZD and a close-in planet. In the case of a planetary Rayleigh scattering atmosphere, small dust particles of the HEZD that also fall in the Rayleigh scattering region contribute significantly less to the total polarized flux than particles of a certain grain radius (see \Cref{masses}). In the case of a planetary Mie scattering atmosphere, however, larger cloud particles severely limit the phase angle range required for the planetary polarized flux to exceed the polarized flux of the dust.

\subsubsection{Potential of analyzing the individual Stokes parameters \(Q\) and \(U\)}\label{Q_U}
\begin{figure}[t!]
\begin{minipage}{0.42\textwidth}
\includegraphics[width=\textwidth]{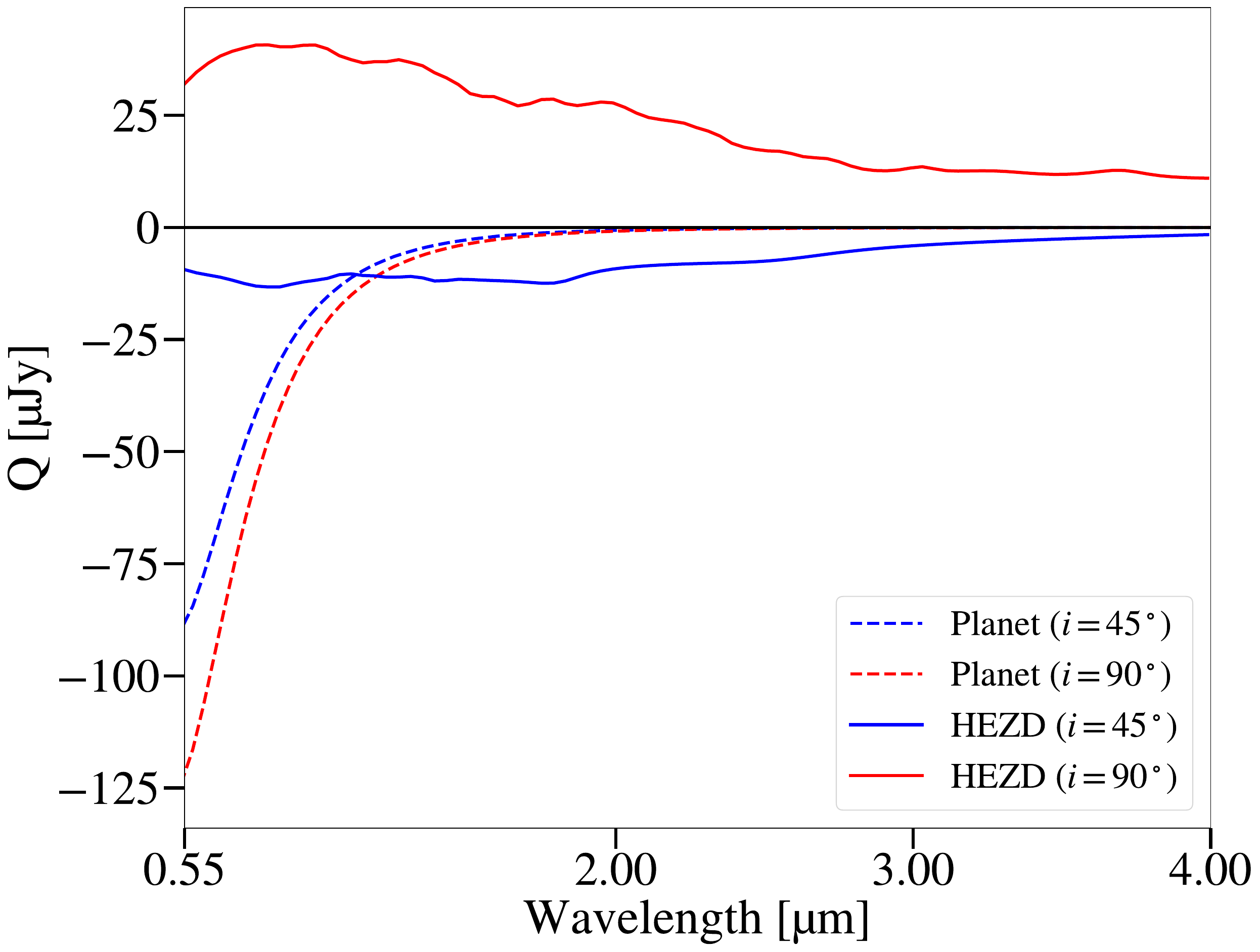}

\end{minipage}
\begin{minipage}{0.42\textwidth}
\includegraphics[width=\textwidth]{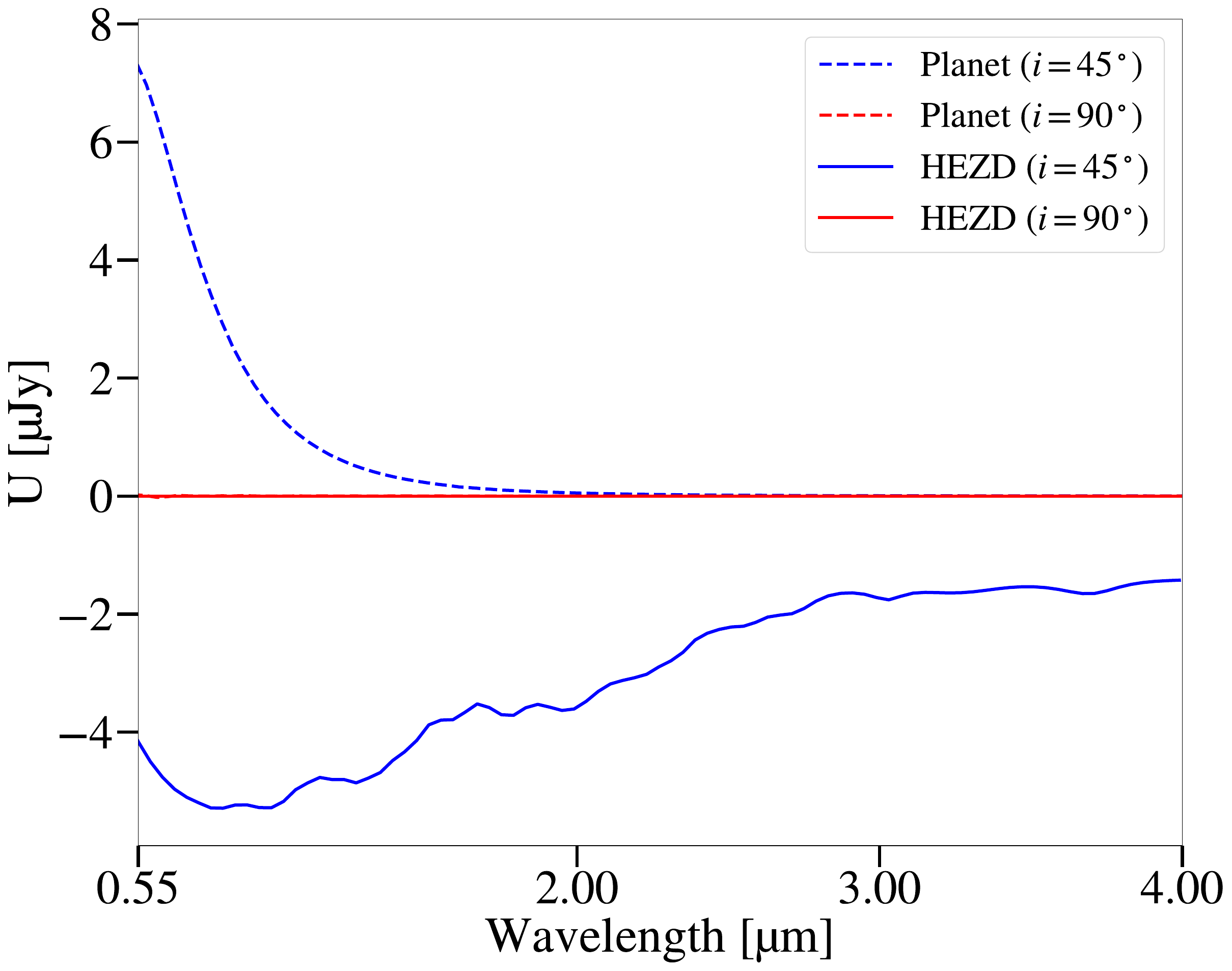}

\end{minipage}
\caption{Wavelength-dependent Stokes parameters for the reference model of the HEZD-planet-star system (see \Cref{table2}) for two orbital inclinations (\(i=45^\circ\), \(90^\circ\)). We show \(Q\) (top) and \(U\) (bottom). See \Cref{Q_U} for details.}\label{Fig 11}
\end{figure}
 
While the degree and orientation of the linear polarization can be derived from the combination of \(Q\) and \(U\), these individual components of the Stokes vector already provide limited information about the polarization state of the radiation. This is made use of in the case of differential polarimetry (e.g., \citealt{Kuhn}; \citealt{Potter}; \citealt{Quanz}). 
Moreover, as illustrated by \citet{Marshall2023}, \(Q\) and \(U\) may provide the basis for distinguishing between the weakly polarized signal of the scientific target (e.g., a debris disk or an exoplanet) from that of the foreground interstellar polarization. 
In the following, we illustrate the potential of the Stokes parameters \(Q\) and \(U\) for the analysis of observations of a system consisting of a star, a HEZD and a hot Jupiter.\\
For this purpose, the two Stokes parameters are shown as a function of wavelength for our reference HEZD and planet seen at two different inclinations (\(i=45^\circ\), \(90^\circ\))  in \Cref{Fig 11}.
Furthermore, the polarizers were assumed to be oriented parallel and perpendicular to the main axis of the disk and the planetary orbit in the case of \(Q\). 
Consequently, \(U\) cancels out in the case of the HEZD ring and for a planetary orbit seen edge-on (\(i=90^\circ\)). Thus, for our defined reference model (edge-on configuration), the total linear polarized flux only results from the Stokes parameter \(Q\). 
We find that  \(Q\) and \(U\) decrease with increasing wavelength. This trend is only slightly modified by the fluctuating wavelength dependence in the case of the Mie scattering-dominated HEZD.\\ 
The generally strong trends of the wavelength dependence on the resulting polarization degree as we found for the planet (strong dependence) and the HEZD (weaker dependence) in \Cref{wavelength-dependence}, is already imprinted in the underlying Stokes parameters \(Q\) and \(U\).
On the one hand, this behavior is due to the decreasing polarization of the radiation of the planet and the HEZD, as illustrated for example in \Cref{Fig 1}.
On the other hand, as both \(Q\) and \(U\) are defined as intensity differences, but for different combinations of the orientation of the polarizers, this behavior is also linked to the decreasing scattered and thus polarized fraction of the radiation of both components (e.g., \Cref{Fig 2}).
In addition to the polarization degree and thus the underlying absolute amount  of both Stokes parameters, the sign of \(Q\) and \(U\) contains additional information about the orientation of the linear polarization. In contrast to the basic analysis of the individual Stokes parameters, it will be particularly difficult to derive the orientation of the linear polarization because it is determined by the ratio of these quantities that is \(\frac{1}{2}\arctan(U/Q)\), resulting in a strong impact of measurement uncertainties. In conclusion, the basic characteristics of the wavelength-dependent polarization state of both the planetary and the HEZD radiation and thus the potential of distinguishing between the two components are already imprinted in the Stokes parameters \(Q\) and \(U\).

\section{Summary and conclusions}\label{Summary}
We compared the wavelength-dependent scattered-light polarization of HEZD and close-in exoplanets starting from a reference model. This model was motivated by observational constraints from the polarization measurements of WASP-18b for a reference close-in planet and a suitable parameter configuration to fit the NIR excess of HD 22484 (10 Tau) as a reference for a HEZD. We varied different model parameters to study their individual impact on the polarization characteristics in the wavelength range from 550 nm to \(4\ \mu\)m. We defined the net polarization degree with and without including the stellar flux separately. The net polarization degree furthermore included the scattered (polarized) radiation of a close-in planet and a HEZD, as well as their thermally reemitted flux. We used a numerical tool (3D Monte Carlo solver POLARIS) for the radiative transfer process in the Rayleigh scattering atmosphere of the close-in planet and an analytical tool to calculate the net scattered-light polarization of the HEZD modeled as an optically thin dust ring. We also investigated the ratio of the the polarized flux of the HEZD to the planetary polarized flux for representative wavelengths of the photometric bands \(V, K\) and \(L\) for selected parameter combinations. We found several trends that have the potential to distinguish the polarization from HEZD and close-in planets. We reach the following conclusions:
\begin{enumerate}

\item The presence of HEZD needs to be considered in any effort to characterize exoplanets via polarimetric observations because they have different wavelength-dependent polarization characteristics and for certain model configurations different orders of magnitude of their scattered-light polarization. This complicates any analysis of the radiation in such a system because the scattered-light polarization of the HEZD may exceed that from the close-in planet for a subset of the parameter space that is usually considered for modeling HEZD. Stricter constraints on the parameters (especially on the grain radii) of HEZD, which can be achieved with interferometers such as MATISSE (\citealt{Lopez}) are thus required to restrict the potentially resulting scattered-light polarization.
\item Within the investigated parameter space, the presence of a HEZD complicates the polarimetric analysis of a close-in planet in the optical wavelength range most under the following conditions: The dust grain radius is 0.1 \(\mu\)m at an observed wavelength of 550 nm, the orbital radius of the HEZD is as small as possible, the HEZD mass is as high as possible, the orbital radius of the planet is as large as possible and the planetary phase angle deviates strongly from  \(67^\circ\).
\item The dust grain radius has the most significant impact on the scattered-light polarization because it strongly changes the wavelength-dependent polarization characteristics of the HEZD and also the total order of magnitude of the polarization (see \Cref{sizes} and \Cref{masses}).
\item The scattered-light polarization of a close-in planet is strongly dependent on the phase angle. For the defined reference planet a phase angle of about \(67^\circ\) results in the highest polarization degree in the \(V\) band (see \Cref{atmosphere} and  \Cref{phase}).
\item The orbital radius of the HEZD, respectively the planet and the HEZD mass are scaling factors that can change the order of magnitude of the polarized flux (see \Cref{masses} and \Cref{phase}).
\item Different compositions of atmospheres of hot Jupiters affect the resulting linear polarization degree only slightly as long as Rayleigh scattering dominates the scattering behavior of the atmospheric and cloud particles. On the other hand, when the absorption of the atmospheric particles is also taken into account and larger cloud particles are considered, then the significant differences between the linear polarization degrees especially in the NIR wavelength range provide a characterization of the atmospheric composition. For cloud particles with sizes comparable to the observing wavelength, Mie scattering dominates. This results in similar wavelength-dependent polarization characteristics for the planet and the HEZD consisting of dust grains of similar radii (the same applies for Rayleigh scattering in the case of small cloud particles and the HEZD consisting of small dust grains, see \Cref{atmosphere} and \Cref{phase}).
\item The potential to distinguish between a HEZD and a close-in exoplanet is already imprinted in the underlying Stokes parameters \(Q\) and \(U\) (see \Cref{Q_U}).
\item Orbital inclinations of the HEZD up to \(90^\circ\) increase the net scattered-light polarization (see \Cref{sizes}).
\item The intrinsic polarization of the stellar radiation of active stars may have a significant impact on the net polarization of the system and thus on the polarimetric analysis (see \Cref{stellar}).
\end{enumerate}
As a next step to complete the investigation of the mutual influence of a planet and a HEZD on their individual polarization signatures, the gravitational impact of the planet on the spatial structure of the HEZD, which might result in asymmetries, has to be considered. The exact mechanisms that sustain or reproduce the HEZD are not yet well understood. The results of such a study would therefore strongly depend on the underlying model and thus on the dominating physical processes. For example, if the HEZD consists of submicron-sized grains close to the star, the dust grains may be trapped in the stellar magnetic field while otherwise, resonant trapping could determine the spatial dust distribution (e.g., \citealt{Kral}).\\ Three of the systems with an observed HEZD from \citet{Kirchschlager2017} also contain exoplanets (\citealt{Pepe}; \citealt{Feng}; \citealt{Lagrange}). They can therefore be selected as a starting point for such an investigation, but these exoplanets are not close-in planets. However, required simulations are difficult to perform because the exact physics providing an explanation of the presence of HEZD first need to be understood. 
\section*{ORCID iDs}
K.Ollmann\ \orcidlink{0009-0003-6954-5252} \url{https://orcid.org/0009-0003-6954-5252}\\
S. Wolf\ \orcidlink{0000-0001-7841-3452} \url{https://orcid.org/0000-0001-7841-3452}\\
M. Lietzow\ \orcidlink{0000-0001-9511-3371} \url{https://orcid.org/0000-0001-9511-3371}\\
 T. A. Stuber\ \orcidlink{0000-0003-2185-0525} \url{https://orcid.org/0000-0003-2185-0525}
\begin{acknowledgements}
This work is supported by the Research Unit FOR 2285 “Debris Disks in Planetary Systems” of the Deutsche Forschungsgemeinschaft (DFG). K.O. and S.W. acknowledge the DFG for financial support under contract WO 857/15-2. We thank the anonymous referee for the very useful suggestions.
\end{acknowledgements}

        %
        %

\begin{appendix}
\section{Model}\label{Model}
At first, we introduce in \Cref{Stokesformalism} the building blocks for calculating the net scattered-light polarization of a HEZD and a close-in planet. Then, we briefly describe the radiation source in \Cref{radiation}, the HEZD model in \Cref{Hotdust} and the planetary model in \Cref{planetary}. This planetary model, which is implemented in the publicly available 3D Monte Carlo radiative transfer code POLARIS (\citealt{Polaris}) and is optimized to handle the required polarization mechanisms in (optically thick) exoplanetary atmospheres, was tested successfully (\citealt{Lietzow, Lietzow2022}).

\subsection{Stokes formalism}\label{Stokesformalism}
We used the wavelength-dependent Stokes parameters to calculate the scattered-light polarization of a HEZD and a close-in planet. They were combined in a 4D vector \(\textbf{S} = (I, Q, U, V)^{T}\) (see e.g., \citealt{Bohren}). For spherical particles the scattering matrix \(\textbf{M}\) has the following structure:
\begin{align} \begin{small}  \label{scattering}\textbf{M}(\em a,\lambda,\theta)=\begin{pmatrix}M_{11}(\em a,\lambda,\theta) &M_{12}(\em a,\lambda,\theta)&0&0\\M_{12}(\em a,\lambda,\theta)&M_{11}(\em a,\lambda,\theta)&0&0\\0&0&M_{33}(\em a,\lambda,\theta)&M_{34}(\em a,\lambda,\theta)\\0&0&-M_{34}(\em a,\lambda,\theta)&M_{33}(\em a,\lambda,\theta)\end{pmatrix},\end{small}   \end{align} with \(\em a\) as the particle radius, \(\lambda\) as the wavelength of the radiation and \(\theta \in [0, \pi]\) as the scattering angle that is the angle between the directions of incident and scattered radiation. The direction of propagation after scattering defines the so-called scattering plane with the original direction of radiation. Before we consider the scattering matrix, the ingoing Stokes vector \(\textbf{S}_{\text{in}}\) has to be rotated around the angle \(\phi \in [0, 2\pi]\) with the rotation matrix \(\textbf{R}(\phi)\) into a different reference frame. Finally, the outgoing Stokes vector \(\textbf{S}_{\text{out}}\) was calculated as \begin{align}\label{outgoing}\textbf{S}_{\text{out}} \propto \textbf{M}\left(\em a,\lambda,\theta\right)\cdot \textbf{R}\left(\phi \right)\cdot \textbf{S}_{\text{in}},  \end{align} 
where the rotation matrix \textbf{R}(\(\phi\)) was set to 
\begin{align}\textbf{R}(\phi)=\begin{pmatrix}1 &0&0&0\\0&\cos(2\phi)&-\sin(2\phi)&0\\0&\sin(2\phi)&\cos(2\phi)&0\\0&0&0&1\end{pmatrix}.\end{align} 

\subsection{Illuminating source}\label{radiation}
As the central illuminating source we assumed a spherical, spatially extended radiation source (star) emitting photon packages. These photon packages are characterized by their point and direction of emission and by their subsequent direction of propagation through the 3D model space. Each package is defined by its wavelength-dependent Stokes vector. The stellar flux that is distributed among the photon packets and measured at the distance to the observer \(d_{\rm{obs}}\) was analytically approximated as 
\begin{align} F_{\star}(\lambda)= \frac{4 \pi^{2} R_{\star}^{2}B_{\nu}\left(T_{\star},\lambda\right)}{4\pi d_{\rm{obs}}^{2}},\end{align}
with \(B_{\nu}\left(T_{\star},\lambda\right)\) as the Planck function in frequency (\(\nu\)) representation for the stellar effective temperature \(T_{\star}\) and the stellar radius \(R_{\star}\). In this context, the resulting fluxes defined in this study were represented by suppressing the index \(\nu\) by the notation \(F_{\nu}(\lambda)=F(\lambda)\).

\subsection{Hot exozodiacal dust}\label{Hotdust}
Because the true nature of HEZD is not known so far, which results in a large possible parameter space (e.g., geometry of the model), we adopted a simplified version of a model that was applied in previous HEZD studies (e.g., \citealt{Kirchschlager2017}). We assumed a ring with orbital radius \(d_{\rm{dust}}\) that consists of spherical graphite dust grains with a single grain radius \(\em a\). Because the HEZD is optically thin, only single-scattering events had to be considered to calculate the resulting (net) polarization. The scattered and polarized flux for a single dust grain measured at the distance to the observer \(d_{\rm{obs}}\) was calculated according to \citet{DMS} as
\begin{equation} \label{singlesca}  F_{\text{sca grain}}(\lambda)=F_{\star}\left(\lambda\right)\frac{\pi \em{a}^{2}Q_{\rm{sca}}\left(\em a,\lambda\right)M_{11}\left(\em a,\lambda,\theta\right)}{4\pi d_{\rm{dust}}^{2}}, \end{equation}
\begin{equation}\label{singlepol} F_{\text{pol grain}}(\lambda)=F_{\star}\left(\lambda\right)\frac{\pi \em{a}^{2}Q_{\rm{sca}}\left(a,\lambda\right)M_{12}\left(\em a, \lambda,\theta\right)}{4\pi d_{\rm{dust}}^{2}}. \end{equation}
The thermal reemission of a dust grain measured at the distance to the observer \(d_{\rm{obs}}\) was defined as
\begin{align}\label{therm}F_{\text{therm grain}}(\lambda)=\frac{4\pi^2 \em{a}^{2}Q_{\rm{abs}}(a,\lambda)\ B_{\nu}\left(T_{\rm{dust}},\lambda\right)}{4\pi d_{\rm{obs}}^{2}}.  \end{align}
The quantities \(Q_{\text{sca}}\) and \(Q_{\text{abs}}\) indicate the efficiency of absorption and scattering. The temperature \(T_{\rm{dust}}\) of a dust grain for a selected radius \(d_{\rm{dust}}\) of the HEZD was calculated with
\begin{equation}\label{dust}d_{\rm{dust}}(T_{\rm{dust}})=\frac{R_{\star}}{2}\sqrt{\frac{\int_{0}^{\infty}Q_{\rm{abs}}\left(\em a,\lambda\right)B_{\nu}\left(T_{\star},\lambda\right)d\lambda}{\int_{0}^{\infty}Q_{\text{abs}}\left(\em a,\lambda\right)B_{\nu}\left(T_{\text{dust}},\lambda\right)d\lambda}}. \end{equation}
The optical properties of the dust grains (\(M_{11},\  M_{12},\ Q_{\text{abs}}\ \text{and}\  Q_{\text{sca}}\)) were computed using the software tool \textit{miex} (\citealt{Wolf}) based on Mie theory (\citealt{Mie}). To calculate the total net scattered-light polarization, we added the scattered (polarized) and thermally reemitted flux for every dust grain according to \Cref{singlesca}, \Cref{singlepol} and \Cref{therm}. We defined the reference plane for the Stokes parameters in the plane in which the dust ring is located. Moreover, we denoted the position vector for the \(n\)-th particle on the dust ring as \(\textbf{r}_{n}\) and the grain position-dependent scattering angle as \(\theta(\textbf{r}_{n})\). We defined the polarized flux of the HEZD for a total HEZD mass of \(\em M_{\rm{dust}}\) distributed among a total number of \(N\) dust grains and for a density \(\rho\) as
\begin{equation}\label{dustflux} F_{\text{pol dust}}(\lambda)=\frac{\em M_{\rm{dust}}}{\rho a^{3}} \sum_{n=1}^{N} F_{\text{pol grain}}(\lambda, \theta(\textbf{r}_{n})).  \end{equation}

\subsection{Planetary atmosphere}\label{planetary}
The structure of the planetary atmosphere was described within a spherical grid. A simple assumption for the radial atmospheric pressure profile was based on the condition of hydrostatic equilibrium and the condition of ideal gas. The wavelength and pressure-dependent absorption and scattering efficiencies of the atmosphere that contribute to the scattering process are given by \citet{Travis} and \citet{Sneep}. In the case of WASP-18b, the pressure presumably ranges from  \(10^{-5}\) bar at the top of the atmosphere to 1 bar (\citealt{Helling}), which defines the lower boundary of the atmosphere. The defined pressure range is thus in the same order of magnitude as in previous radiative transfer simulations for hot Jupiters (e.g., \citealt{Bailey2018}). In addition to gaseous particles, we also considered clouds located within \(10^{-2}\ \mathrm{bar} -10^{-4}\) bar with \(n(r)\) as a power-law size distribution of particles with radius \(r\), described by \citet{Hansen} as \begin{equation}\label{clouds} n(r) \propto r^{(1-3\nu_{\rm{eff}})/\nu_{\rm{eff}}} e^{-r/(r_{\rm{eff}}\nu_{\rm{eff})}},\end{equation} with \(r_{\rm{eff}}\) as effective particle radius and \(\nu_{\rm{eff}}\) as effective variance. Additionally, their optical depth was defined to be one at a wavelength of 550 nm. \\
After emission from the stellar photosphere a photon package travels toward the planet with radius \(R_{\text{planet}}\), it is scattered by atmospheric particles and finally detected by an observer. The planetary distance to the illuminating source (orbital radius) is given by \(d_{\text{planet}}\) and the angle between the illuminating source and the observer as seen from an illuminated surface is the phase angle \(\alpha\) (\(\alpha=\pi-\theta\) for single scattering). We applied POLARIS to calculate the resulting Stokes vectors  \(I\), \(Q\) and \(U\), and the planetary scattered and polarized flux for selected \(\alpha\) observed at \(d_{\rm{obs}}\) were
\begin{equation}\label{fluxplanetsca}F_{\text{sca planet}}(\lambda)=I,\end{equation} 
\begin{equation}\label{fluxplanet}F_{\text{pol planet}}(\lambda)=\sqrt{Q^{2}+ U^{2}}.\end{equation}
The planetary thermally reemitted flux observed at \(d_{\rm{obs}}\) was analytically approximated as
\begin{equation}\label{planetflux} F_{\text{therm planet}}(\lambda)=\frac{4\pi^{2} R_{\text{planet}}^{2} B_{\nu}\left(T_{\text{planet}},\lambda\right)}{4\pi d_{\rm{obs}}^{2}}. \end{equation} 
For this study we neglected that \(F_{\text{therm planet}}\) can also be scattered in the planetary atmosphere and thus also be polarized in the case of a flattened planet or a inhomogeneous cloud distribution (\citealt{De Kok}). The associated planetary equilibrium temperature \(T_{\text{planet}}\) was determined by
\begin{equation}\label{Teq}T_{\rm{planet}}=\left(\frac{R_{\star}^{2}T_{\star}^{4}}{4d_{\text{planet}}^{2}}\right)^{1/4}.\end{equation}
\section{Further case studies}\label{Addition}
\subsection{Spectral classes}\label{Spectral}
\begin{figure}[t!]
\includegraphics[scale=0.22]{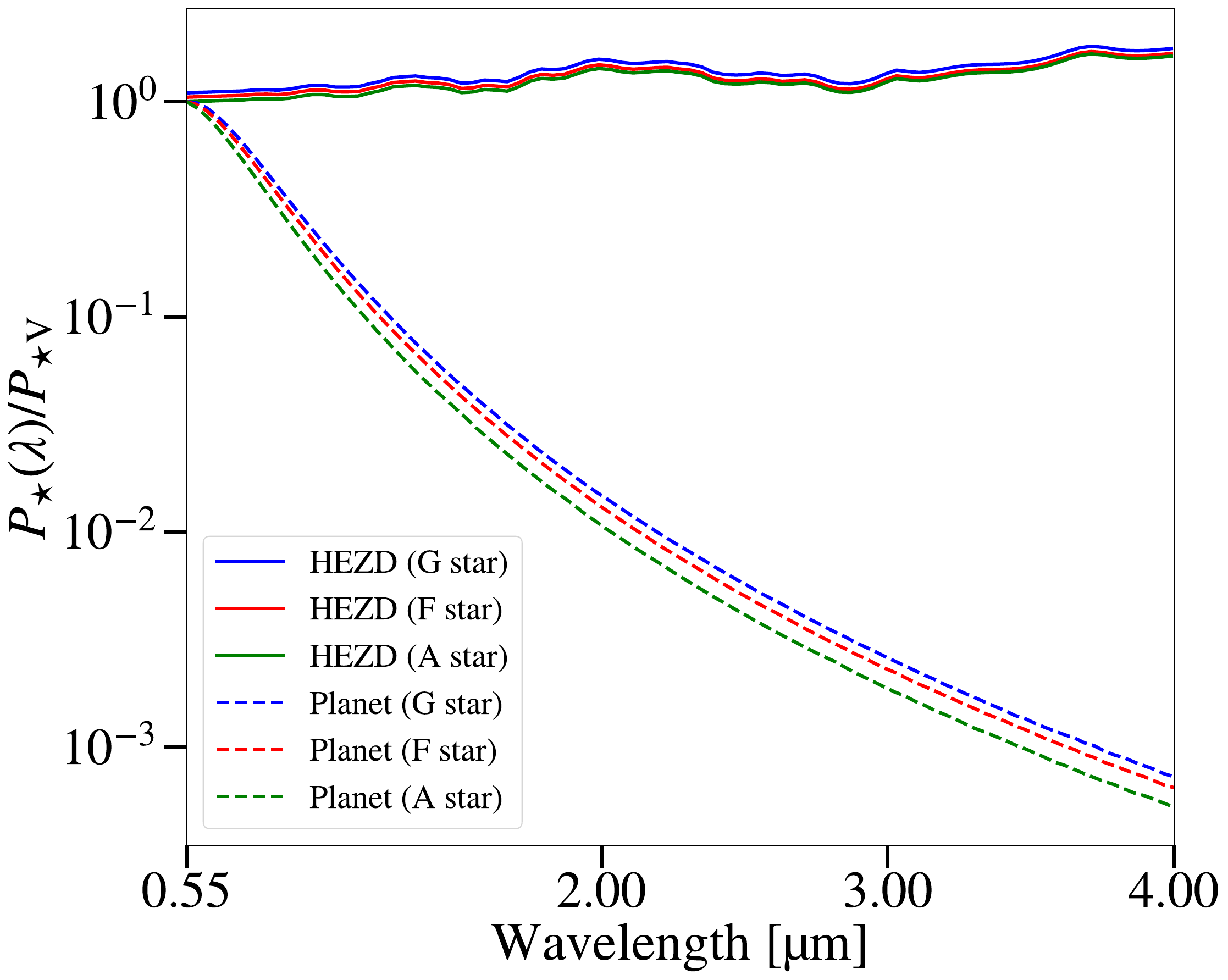}
\caption{Polarization as a function of wavelength relative to the polarization at a wavelength of 550 nm for G, F and A stars for the reference model of the HEZD and the planet. See \Cref{Spectral} for details.} \label{Fig 12}
\end{figure}
Because HEZD and close-in planets have not only been detected around unpolarized F stars that were considered in the main part of this study but also, for example, around G and A stars (e.g., \citealt{Absil2013}), we additionally investigated the influence of a G and A star ($\tau  \ \mathrm{Cet}$: \(T_{\star}=\) 5290 K, \(R_{\star}=0.7\ R_{\odot}\);\:\: $\alpha\ \mathrm{Lyr}$: \(T_{\star}=\) 9620 K, \(R_{\star}=2.22\ R_{\odot}\), stellar radii and effective temperatures from \citet{Kirchschlager2017}) on the wavelength-dependent polarization characteristics of a HEZD and a close-in planet (see \Cref{Fig 12}). When compared to the trend for the F-type reference star, we find that the planetary polarization decrease slightly steeper for an increasing wavelength in case of the A star than  for the F or G star. However, the differences in the case of the polarization degree of the HEZD are even smaller (\(< 5\%\)). We therefore conclude that the spectral class of the star has no significant influence on the wavelength-dependent polarization characteristics of a close-in planet and a HEZD.
\subsection{Distribution of the dust grain size}\label{distribution}
\begin{figure}[t!]
\includegraphics[scale=0.25]{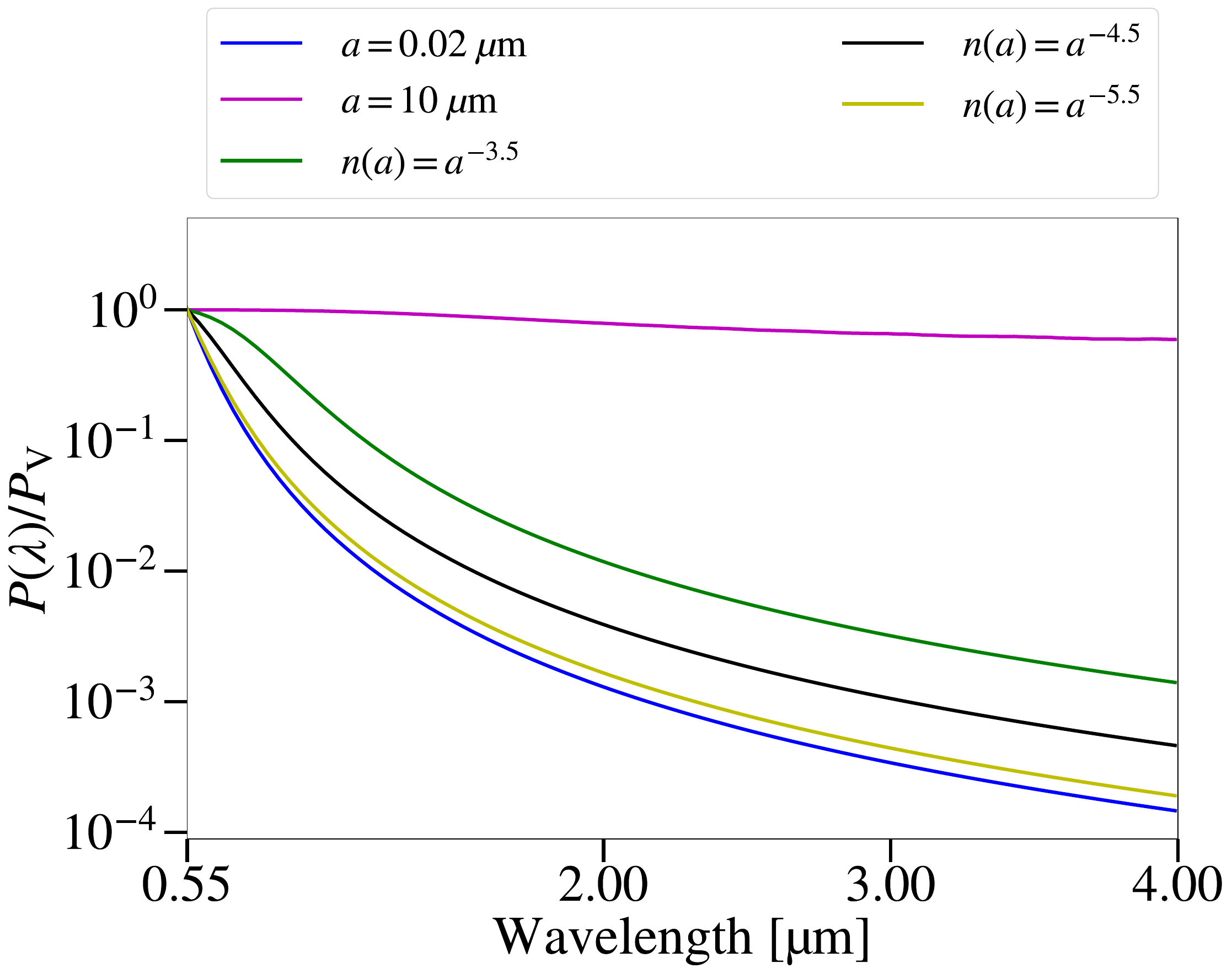}
\caption{Wavelength-dependent polarization of the HEZD relative to the corresponding polarization at a wavelength of 550 nm for two selected dust grain radii 
and three selected dust grain size distributions. See \Cref{distribution} for details.} \label{Fig 13}
\end{figure}
 
Although the observed wavelength-dependence of the NIR excess of a HEZD can be sufficiently well reproduced by a narrow ring consisting of dust grains with a single dust grain size \citep{Kirchschlager2017,Kirchschlager2018}, the lack of detections at longer wavelengths does not allow us to exclude steep dust grain size distributions that also contain a small fraction of significantly larger particles  (e.g., \citealt{Lebreton}).\\ For this reason, dust grain size distributions with grain radii ranging from \(0.02\ \mu\)m to \(10\ \mu\)m are considered in the following discussion of the impact of a dust grain size distribution in contrast to that of dust grains with a single size (see \Cref{Fig 13}). We assumed a power-law grain size distribution \(n(\em a)=a^{\gamma}\) for three selected values of \(\gamma\) (\(-3.5,-4.5, -5.5\)). We find two important trends: First, for a single dust grain radius of \(0.02\ \mu\)m (dominating scattering mechanism: Rayleigh scattering) the dust polarization ratio decreases steeply with increasing wavelength, while for a single dust grain radius of \(10\ \mu\)m (dominating scattering mechanism: Mie scattering) there is almost no decrease at all (see \Cref{sizes} for a more detailed discussion of the influence of a single dust grain size on the polarization of HEZD). Second, considering a grain size distribution, we find that the decrease in the exponent of the grain size distribution (\(\gamma=-3.5 \rightarrow -5.5\)) results in a polarization ratio that approaches the ratio we found for the smallest single dust grain size (\(0.02\ \mu\)m). This finding is expected because for steep size distributions the optical properties and thus the resulting polarization characteristics of the HEZD are determined almost exclusively by the smallest dust grains. The existing observations of the HEZD phenomenon favor a dominance of dust grains in the submicron range and thus HEZD models with a steep (\(\gamma<-3.5\)) power-law grain size distribution.
\end{appendix}
\end{document}